\documentstyle[12pt]{article}

\let\ph=\varphi

\def\0{\over } \def\1{\vec }     \def\2{{1\over2}} \def\4{{1\over4}}
\def\5{\bar }  \def\6{\partial } \def\7#1{{#1}\llap{/}}
\def\8#1{{\textstyle{#1}}}       \def\9#1{{\bf {#1}}}
 \def\llp{\hbox to 0pt{\hss /\hskip1.5pt}}
\def\llo{\hbox to 0.2pt{\hss /}} \def\llq{\hbox to 0pt{\hss /\hskip0.5pt}}
\def\so{\supset\hbox to 0pt{\hss $\displaystyle -$\hskip1pt}}

\def\<{\langle } \def\>{\rangle }

\let\nn=\nonumber
\def\bea{\begin{eqnarray}} \def\eea{\end{eqnarray}}
\def\beann{\begin{eqnarray*}} \def\eeann{\end{eqnarray*}}
\def\beq{\begin{equation}} \def\eeq{\end{equation}}

\newcommand{\vp}{\varphi}
\newcommand{\prop}{\bigtriangleup}
\date{}
\title{
{\large\rm DESY 93-147}\hfill{\large\tt ISSN 0418-9833}\\
{\large\rm November 1993}\hfill\vspace*{3.5cm}\\
Aspects of the Cosmological Electroweak \\
Phase Transition}
\author{
D. B\"odeker$^a$, W. Buchm\"uller$^b$, Z. Fodor$^b$\thanks{On leave from
Institute for Theoretical Physics, E\"otv\"os
University, Budapest, Hungary}  and T. Helbig$^b$\\
\\
\normalsize\it a II. Institut f\"ur Theoretische Physik, Universit\"at
Hamburg, Germany\\
\normalsize\it b Deutsches Elektronen-Synchrotron DESY, Hamburg, Germany
\\
}
\begin{document}
\setlength{\baselineskip}{18pt}
\maketitle
\vspace*{1cm}
\begin{abstract}
We study the decay of the metastable symmetric phase in the standard
model at finite temperature. For the SU(2)-Higgs model
the two wave function correction terms $Z_{\vp}(\vp^2,T)$
and $Z_{\chi}(\vp^2,T)$ of Higgs and Goldstone boson fields
are calculated to one-loop order. We find that the derivative
expansion of the effective action
is reliable for Higgs masses smaller than the W-boson mass.
We propose a new procedure to evaluate the
decay rate by first integrating out the vector field
and the components of the scalar fields with non-zero Matsubara
frequencies. The static part of the scalar field is treated in the saddle point
approximation. As a by-product we obtain a formula for the decay rate of
a homogeneous unstable state. The course of the cosmological
electroweak phase transition is evaluated numerically for different Higgs
boson masses and non-vanishing magnetic mass of the gauge boson. For Higgs
masses above $\sim 60$ GeV the latent heat can reheat the system to the
critical temperature,
which qualitatively changes the nature of the transition.
\end{abstract}
\newpage
\section{Introduction}
At high temperatures the spontaneously broken symmetry of electroweak
interactions is restored \cite{kirlin1}-\cite{kirlin2}. This is a direct
consequence of the Higgs mechanism of symmetry breaking, and the
corresponding phase transition is an intriguing aspect of electroweak
interactions. It is also of great cosmological importance since at
temperatures of order the critical temperature of the phase transition
baryon-number violating processes fall out of thermal equilibrium
\cite{rubetal}.
As a result, the present cosmological baryon asymmetry is finally
determined at the electroweak phase transition.

Theoretical descriptions of the phase transition start from the
finite-temperature effective potential whose local minima yield the
free energy of states with a homogeneous Higgs field given by the
position of the local minimum (cf.\ \cite{kapusta}). Considerable
effort has been devoted to evaluate the effective potential in
perturbation theory \cite{dine}-\cite{zwirner} where, due to infrared
divergences, a useful expansion is only obtained after a
resummation of an infinite number of terms. Alternatively, interesting
first results have been obtained by lattice Monte Carlo simulations
\cite{jersak}-\cite{rummu}, reduced three-dimensional
action \cite{patkos} and by using an average action at
finite temperature \cite{wetterich}.

The decay of a metastable homogeneous state involves ``critical droplets''
and therefore inhomogeneous field configurations. The free energy of
these configurations is usually estimated based on an effective action
which is the sum of the finite-temperature effective potential and
the canonical kinetic term of the Higgs field. We will improve this
approximation by evaluating the finite-temperature
wave function correction terms of the SU(2)-Higgs model to one-loop
order. This will allow us to determine the range of Higgs boson masses
for which the derivative expansion of the effective action is reliable.

Following Langer's theory of metastability \cite{langer}
we evaluate the decay rate in
the semiclassical approximation where one starts from some approximate
``coarse-grained'' effective action and expands around its saddle point
interpolating between the symmetric and the broken phase.
For finite-temperature quantum field theories
the ``coarse-grained'' effective action is frequently approximated
by the one-loop effective potential
together with canonical kinetic terms for the scalar fields.
The
Callan-Coleman theory for the decay of the false vacuum at zero temperature
in four dimensions \cite{callan} is then applied to the effective
three-dimensional theory at finite temperature \cite{linde}.

One problem, which points to some inconsistency of this approximation,
is that the occuring effective potential is in general complex.
The origin of this problem in the calculation of the decay rate
is the incorrect treatment of the scalar fluctuations which
appear twice, as contribution to the effective potential
and as fluctuations around the saddle point. Instead, one has to split
the functional integration into two parts, such that the first one yields
a ``coarse-grained'' effective action which has a barrier between the
symmetric and the broken phase. In the second part a perturbative
expansion is then performed around the non-trivial saddle point,
and only here long wave length scalar fluctuations appear.
In this
spirit, in \cite{bhw} the integration over the vector fields has been
performed in the first step. The same approach has been followed in
\cite{eweinberg} for the analogous problem of the Coleman-Weinberg model at
zero temperature, in cite{bfhw} for nonabelian gauge theories and
in \cite{gleiser} for Yukawa models at finite temperature.
Here we will further improve these
calculations by integrating in the first step also over the components
of the scalar fields with non-zero Matsubara frequencies. This will
lead us to a ``coarse-grained'' effective action
which is similar to, but not identical with
the one used by Dine et al.\ \cite{dine}.

Several aspects of the cosmological electroweak phase transition, such
as nucleation and growth of critical droplets, wall velocity, transition
time etc.\ have already been discussed in the literature (cf.\ \cite{dine},
\cite{bfhw},\cite{enqvist}-\cite{carkap}).
We shall use our results on the finite-temperature effective action
to study in more detail the dependence of the course of the phase transition
on the Higgs boson mass and also the influence of a non-zero magnetic mass
of the W-boson. Our numerical calculations show that reheating effects
can qualitatively change the nature of the transition for Higgs masses
larger than $\sim 60$ GeV.

The paper is organized as follows. In sect.\ \ref{finitet} we evaluate the
two wave function correction factors for the SU(2)-Higgs model to
one-loop order. The decay of metastable and unstable homogeneous states
is discussed in sect.\ \ref{falsev}. The course of the phase transition
is studied analytically and also in detail numerically
in sect.\ \ref{cosmopt}. The main results are summarized in sect.\ 5.
\section{The effective action at finite temperature}\label{finitet}
Consider the SU(2)-Higgs model at finite temperature which is described
by the action
\beq
\label{model}
S_{\beta}[\Phi,W] = \int_{\beta}dx\left(
-{1\over 4}W^a_{\mu\nu}W^{a\mu\nu} + (D_{\mu}\Phi)^{\dagger}(D^{\mu}\Phi)
- V_0(\vp^2)\right) \quad,
\eeq
where
\bea
V_0(\vp^2)={1\over2}\mu^2\vp^2+{1\over4}\lambda\vp^4\ ,\quad
{1\over 2}\vp^2 \equiv \Phi^{\dagger}\Phi \quad, \nn \\
\int_{\beta}dx\equiv\int_0^{\beta}d\tau \int d^3x \quad ,\quad
\beta\equiv {1\over T}.
\eea
$D_{\mu}$ and $W^a_{\mu\nu}$ are the covariant derivative and the Yang-Mills
field strength, respectively.
The gauge fixing part of the lagrangian has
been omitted, and we will always work in Landau gauge.
The SU(2)-Higgs model contains the essential features
of the standard model of electroweak interactions. Setting
$\sin{\theta_W}=0$, one neglects the mass difference between
Z- and W-boson relative to the W-boson mass. This corresponds to
the present accuracy in the description of the electroweak phase
transition. We include the effects of the three generations of
quarks and leptons with the usual gauge and Yukawa couplings.

In order to find the thermal equilibrium states and to compute properties
of the first-order phase transition we have to know the effective action
$\Gamma_{\beta}[\Phi]$
of the theory at finite temperature as functional of the
Higgs field $\Phi$ which, after gauge fixing, plays the role of
an order parameter. The effective action can be expressed as a
functional integral
\beq \label{effaction}
e^{\textstyle -\Gamma_{\beta}[\Phi]} =
\int_{\beta}[D\hat{\Phi}][D\hat{W}_{\mu}]\exp{
\left(-S_{\beta}[\Phi+\hat{\Phi},\hat{W}]
+\int_{\beta}dx\frac{\delta\Gamma_{\beta}[\Phi]}{\delta\Phi(x)}
\hat{\Phi}(x)\right)}\quad,
\eeq
where the integration extends over fields with periodic boundary
conditions in $\tau$. Again, gauge fixing and ghost terms have not been
written explicitly.
For stationary, $\tau$-independent fields
$\Phi(\vec{x})$
the effective action corresponds to the free energy density of the system,
\beq
\Gamma_{\beta}[\Phi] = \beta F[\Phi,T] \quad.
\eeq
Ordinary perturbation theory to any finite order
does not yield a good approximation to
$\Gamma_{\beta}[\Phi]$ as
the scalar masses are negative for small values of $\vp$, and because of
the infrared behaviour of the finite-temperature perturbation series.

A systematic expansion in the gauge coupling $g$ and the scalar
self-coupling $\lambda$ is obtained by means of a resummation, where
in all propagators the tree-level
masses are replaced by one-loop plasma masses to order $g^2$ and $\lambda$
(cf. \ \cite{bfhw}). The mass differences are then treated as counter terms,
\beq \label{counter}
\delta S_{\beta} = -{1\over2}\int_{\beta}dx\left[W^{a\mu}\left(
\delta m^2_L P_{L\mu\nu} + \delta m_T^2 P_{T\mu\nu}\right)W^{a\nu}
-\delta m^2_{\vp} \ph^2 -\delta m^2_{\chi} \chi^2 \right] \quad.
\eeq
Here $P_L$ and $P_T$ are the usual projection operators
on longitudinal and transverse degrees of freedom of the vector field
(cf.\ \cite{kapusta}).
The plasma mass terms $\delta m^2_i$ are the differences between the
masses
\bea
m_L^2&=&{11\over 6}g^2 T^2+{g^2\vp^2 \over 4} \quad, \label{ml}\\
m_T^2&=&{g^2\vp^2 \over 4} \quad, \label{mt}\\
m_\vp^2&=&\left({3\over 16}g^2+{\lambda\over 2}+{1 \over 4}f_t^2\right)
(T^2-T_b^2)+3\lambda\vp^2 \quad,\label{mvp}\\
m_\chi^2&=&\left({3\over 16}g^2+{\lambda\over 2}+{1 \over 4}f_t^2\right)
(T^2-T_b^2)+\lambda\vp^2 \label{mchi}\quad,
\eea
and the tree-level masses
\beq
m^2={1\over 4}g^2 \vp^2 \quad,\quad
\bar{m}^2_{\vp}=\lambda(3\vp^2-v^2) \quad,\quad
\bar{m}^2_{\chi}=\lambda(\vp^2-v^2) \quad.
\eeq
In the plasma masses we have included the effect of the top-quark loop, with
$m_t=f_t v /\sqrt{2}$. The temperature $T_b$, at which the barrier
between the symmetric and the broken phase vanishes, reads in
leading order of the couplings
\beq\label{Tb}
T_b^2={16\lambda v^2 \over 3g^2+8\lambda +4f_t^2} \quad.
\eeq
Note, that the tree-level Higgs- and Goldstone-boson masses are obtained
from the scalar potential by ($\vp^2=\sum_{I=1}^4 \vp_I\vp_I$)
\beq
\frac{\partial^2 V_0}{\partial\vp_I\partial\vp_J}=
{\bar {m}}^2_\vp P^\vp_{IJ} + {\bar {m}}^2_\chi P^\chi_{IJ} \quad,
\eeq
where
\beq
P^\vp_{IJ}=\frac{\vp_I\vp_J}{\vp^2} \quad,\quad
P^\chi_{IJ}=\delta_{IJ}-\frac{\vp_I\vp_J}{\vp^2}
\eeq
are the projection operators on the Higgs and the Goldstone fields.

At one-loop order the improved perturbation theory yields
the effective potential to order $g^3,\ \lambda^{3/2}$,
\bea \label{thepotential}
V_{eff}(\vp^2,T)&=&{1 \over 2}\left({3g^2\over 16}+{\lambda\over 2}
+{1 \over 4}f_t^2\right)
(T^2-T_b^2)\vp^2+{\lambda\over 4}\vp^4  \nonumber\\
        & &-(3 m_L^3+6 m_T^3+m_\vp^3+3 m_\chi^3)
\frac{T}{12\pi}+{\cal O}(g^4,\lambda^2,f_t^4),
\eea
which is equivalent to the result of the ring summation \cite{carrington}.
Several two-loop contributions are also known.
For the SU(2)-Higgs model the effective potential has been calculated
to order $g^4,\ \lambda$ \cite{arnold},
and for the abelian Higgs model it is known
to order $e^4,\ \lambda^2$ \cite{hebecker}.

The strength of the electroweak phase transition is rather sensitive to the
magnetic mass of the gauge bosons, whose calculation requires non-perturbative
techniques. In Landau gauge the one-loop gap equations yield
$m_T=(g^2/3\pi)T$ at $\vp=0$ \cite{bfhw},\cite{zwirner}.
In order to estimate its effect
on parameters of the phase transition we will replace eq. (\ref{mt})
by
\beq \label{truemt}
m_T^2 = \gamma^2 {g^4\over 9\pi^2} T^2 + {g^2\over 4}\vp^2 \quad,
\eeq
and compute sensitive quantities for different values of $\gamma$.

The effective potential (\ref{thepotential}) has degenerate
local minima at $\vp=0$ and $\vp=\vp_c>0$
at a critical temperature $T_c$ and
therefore predicts a first-order phase transition.
The evaluation of the transition
rate requires knowledge of a stationary point of the free energy which
interpolates between the two local minima.
The effective action (\ref{effaction})
can be expanded in powers of derivatives, and for time-independent fields
one has
\beq \label{expander}
\Gamma_{\beta}[\Phi] = \beta \int d^3x \left( V_{eff}(\vp^2,T)
+ {1\over2} (\delta_{IJ} + Z_{IJ}(\Phi,T))
\vec{\nabla}\vp_I\vec{\nabla}\vp_J
+ \ldots\right)\quad.
\eeq
{}From eqs.\ (\ref{effaction}) and (\ref{expander}) it is obvious that the
functions $Z_{IJ}$ can be obtained from the inverse scalar propagator
in the homogeneous scalar background field $\Phi$
with spacial momentum $\vec{k}$,
\bea
D^{-1}(\vec{k}^2,\Phi,T)_{IJ} &=&
\vec{k}^2\delta_{IJ}+\frac{\partial^2 V_0(\vp^2)}{\partial\vp_I\partial\vp_J}
+i\Sigma_{IJ}(\vec{k}^2,\Phi,T) \nn\\
&=&m^2_{\vp}(\vp^2,T)P^{\vp}_{IJ}+m_{\chi}^2 (\vp^2,T) P^{\chi}_{IJ}\nn\\
& &+ (\delta_{IJ} + Z_{IJ}(\Phi,T))\vec{k}^2 + \cal{O}(\vec{k}^4)\quad.
\eea
The one-loop contributions to $Z_{IJ}$ are shown in fig.\ (\ref{zfac}).
It is a well-known problem in finite temperature field theory that the
order of a given Feynman graph can be smaller than the order coming from the
vertices.  This is due to the fact that loop diagrams accumulate couplings
in the denominators which decrease the power of the usual couplings coming
from the Feynman rules. In appendix A we show that only graphs of the type
shown in fig. 1 give contributions to the wave function correction
in leading order,
where one has to include not only the tree level but also the plasma masses
for internal lines. Note, that the knowledge of the effective action
for static fields is not sufficient if one is interested in time-dependent
processes such as the thermalization of perturbations \cite{vilja}.

Eq.\ (\ref{expander}) is an expansion in powers of $(\vec{\nabla}/m_i)^2$,
where $m_i$ denotes all masses which appear in propagators of internal
lines. Clearly, the terms with zero Matsubara frequencies for the
internal lines dominate, since they have the smallest denominators.
For the same
reason the top-quark contribution is suppressed, since all Matsubara
frequencies are non-zero for fermions. It is straightforward to
evaluate the contributions of fig.\ (\ref{zfac}) as function of the couplings
$g$ and $\lambda$, the scalar propagators
\beq
\prop(\vec{k}^2,m^2_i)={1\over \vec{k}^2+m^2_i} \quad,
\eeq
and the vector propagator
\beq
D_{\mu\nu}(\vec{k})=\prop(\vec{k}^2,m_L^2)P_{L\mu\nu}
+\prop(\vec{k}^2,m_T^2)P_{T\mu\nu} \quad.
\eeq
After some algebra one obtains a sum of integrals of the type
\bea
I(\vec{k}^2,m_1^2,m_2^2) &=&
T\int\frac{d^3l}{(2\pi)^3}\prop(\vec{l}^2,m_1^2)
\prop((\vec{l}+\vec{k})^2,m_2^2) \\
&=& \frac{T}{4\pi\mid\vec{k}\mid}\arctan{\frac{\mid\vec{k}\mid}{m_1+m_2}} \\
&=&{1\over4\pi}\frac{T}{m_1+m_2}\left(1-{1\over3}\frac{\vec{k}^2}
{(m_1+m_2)^2} + \cal{O}(\vec{k}^4)\right)\quad.
\eea
Collecting all terms we obtain the final result
\beq \label{ztot}
Z_{IJ}(\Phi,T) = Z_{\vp}(\vp^2,T)P^{\vp}_{IJ}+Z_{\chi}(\vp^2,T)P^{\chi}_{IJ}
\quad,
\eeq
where
\bea \label{zvp}
Z_{\vp}(\vp^2,T) &=& {T\over4\pi}\Bigg[{1\over4}\lambda\bar{m}^2
\left({3\over m_{\vp}^3}+{1\over m_{\chi}^3}\right)
-2 g^2 {1\over m_{\chi}+m_T}  \nn \\
 & &\qquad +{1\over8}g^2 m^2
\left({1\over 2m_L^3}+{5\over m_T^3} \right)\Bigg]
\eea
and
\beq \label{zchi}
Z_{\chi}(\vp^2,T)= {T\over 6\pi}\left[\frac{2\lambda \bar{m}^2}
{(m_{\vp}+m_{\chi})^3}-g^2\left({2\over m_{\chi}+m_T} + {1\over m_{\vp}+m_T}
\right)\right] \quad,
\eeq
with $\bar{m}^2=\lambda \vp^2$ and $m^2=g^2\vp^2/4$.
In terms of the complex field $\Phi$ this
result can be written as
\beq
{1\over2}Z_{IJ}\vec{\nabla}\vp_I\vec{\nabla}\vp_J =
{1\over4} (\Phi^{\dagger}\Phi)^{-1}
[\vec{\nabla}(\Phi^{\dagger}\Phi)]^2 Z_{\vp}
+\left(\vec{\nabla}\Phi^{\dagger}\vec{\nabla}\Phi
-{1\over4} (\Phi^{\dagger}\Phi)^{-1}
[\vec{\nabla}(\Phi^{\dagger}\Phi)]^2 \right)Z_{\chi} \quad.
\eeq
(For completeness we give in appendix A the wave function correction
term $Z(\varphi^2,T)$ for the abelian Higgs model.)
The $Z$-factor $Z_{\vp}$ for the Higgs field
is dominated by the contribution from the transverse
vector boson loop which, if one neglects the magnetic mass, diverges at
$\vp \sim 0$,
\beq \label{znew}
Z_{\vp}(\vp^2,T) \sim {5g^2\over16\pi}{T\over \vp} \quad.
\eeq
Neglecting plasma mass terms for the vector bosons
and using $m_L = m_T \gg m_{\chi}$
one obtains instead
\beq \label{zmoss}
Z_{\vp}(\vp^2,T) \sim -{21 g^2\over32\pi}{T\over \vp} \quad.
\eeq
This  agrees with the result of a previous calculation by
Moss et al.\ \cite{moss} who used a different technique.
The comparison of eqs.\ (\ref{znew}) and (\ref{zmoss}) shows the
important difference between the ordinary and the improved perturbation
theory with plasma masses\footnote{In a very recent paper \cite{jung}
$Z_{\vp}$ has also been calculated using the improved perturbation
theory. Our results differ from those obtained by these authors.
The differences can be traced back to the treatment of the mass
counter terms of eq.\ (\ref{counter}).}.
The order of magnitude for $Z_{\vp}$ is the
same in both cases, the sign, however, has changed.
Wave function correction terms in the 't Hooft-Feynman background gauge
have recently been evaluated in \cite{schmidt}.
Note, that despite the divergence of $Z_{\vp}$ at $\vp \sim 0$ the
correction to the surface tension
$\sigma = \int_0^{\vp_c}d\vp\sqrt{2(1+Z_{\vp}(\vp^2,T_c))V_{eff}(\vp^2,T_c)}$
is finite.

What is the effect of the wave function factors $Z(\vp^2,T)$ on parameters
describing the electroweak phase transition?
The first-order phase transition is due to the barrier in the effective
potential (\ref{thepotential}) which is
essentially generated by the vector-boson
loops. This implies that at the critical temperature $T_c$ the magnitude
of the local minimum in the broken phase is $\vp_c/T_c \ \sim\ g^3/\lambda$.
One also easily verifies that $(T_c^2-T_b^2)/T_c^2\ \sim\ g^4/\lambda$.
Hence, in the effective potential (\ref{thepotential}) the terms quadratic,
cubic and quartic in $\vp$ are all of order $\sim (g^{12}/\lambda^3) T_c^4$.
The perturbative approach is only reliable if this energy density is larger
than the density due to the magnetic mass of the W-boson, which is of
order $m_T^3 T_c\ \sim\ g^6 T_c^4$. Hence a necessary condition for
the smallness of non-perturbative effects is  $\lambda < g^2$.
Further, if one uses the high-temperature expansion one has to satisfy
$g\vp_c/T_c < 1$, i.e., $\lambda > g^4$. In this case the allowed
range of $\lambda$ is given by
$g^4\, <\, \lambda\, < \, g^2
$.

A naive estimate of the size of $Z(\vp^2_c, T_c)$ can now be
obtained based on eqs. (\ref{ml}) - (\ref{mchi}), (\ref{zvp}) and (\ref{zchi}).
One easily verifies that, at
$\vp_c$ and $T_c$, the plasma masses are of order
\beq
m_T\sim {g^4\over\lambda}T_c\,<\, m_L\, \quad,\quad
m_{\vp,\chi}\sim {g^3\over\sqrt{\lambda}}T_c \quad.
\eeq
{}From eqs. (\ref{zvp}), (\ref{zchi}) one then concludes
\beq
Z_{\vp,\chi}\,=\,\cal{O}\left({\lambda\over g^2}\right)\quad.
\eeq
Hence, for the allowed range of couplings determined from the effective
potential corrections due to the $Z$-factors can be treated as perturbation.

Clearly, these estimates are rather naive. As the saddle point of the effective
action, which interpolates between the two local minima, varies from
$\vp=0$ to $\vp=\vp_c$, the full scalar mass terms
\beq \label{truemasses}
m^2_{\vp}=\frac{\partial^2 V_{eff}(\vp^2,T)}{\partial \vp^2}\,  ,\quad
m^2_{\chi}={1\over\vp}\frac{\partial V_{eff}(\vp^2,T)}{\partial \vp}
\eeq
become negative and the expansion in powers of derivatives breaks
down.
Even for positive mass terms this expansion is problematic for
stationary points of the effective action. For these field
configurations
\begin{eqnarray}\label{eqmot}
{\vec\nabla}^2 \varphi&=&
\frac{\partial}{\partial\varphi}\,V_{eff}(\varphi^2,T)\nonumber\\
&=&m_\chi^2\varphi.
\end{eqnarray}
the equation of motion shows that $\vec{\nabla}/m_i={\cal {O}}(1)$.
Hence, higher order terms in the derivative expansion are
non-negligible
As we shall see in the following section, a more
careful treatment nevertheless confirms the above conclusion
on the size of the wave function correction terms.

\section{Decay of the false vacuum} \label{falsev}
At temperatures below the critical temperature $T_c$ the symmetric
phase becomes metastable and decays via nucleation and growth of
droplets of the broken phase. In condensed matter physics a theory
for the decay of such metastable states has been developed by Langer
many years ago \cite{langer}. Starting from the Fokker-Planck equation
for the probability distribution of droplets with different sizes, Langer
derived an expression for the nucleation rate in terms of a saddle-point
$\bar{\vp}$
of the free energy which interpolates between the two phases. A similar
result for the decay rate has been obtained by applying the
Callan-Coleman theory for the decay of the false vacuum at zero temperature
in four dimensions \cite{callan} to the effective
three-dimensional theory at finite temperature \cite{linde}. We will
base our discussion on Langer's theory according to which the decay rate
is given by
\beq\label{decay}
\Gamma={\kappa\over 2\pi}\,\frac{\mbox{Im} Z_\beta
[\bar{\Phi}]}{Z_\beta[\Phi=0]} \quad,\quad
Z_\beta[\Phi] = e^{\textstyle -\Gamma_{\beta}[\Phi]}  \quad,
\eeq
where $\Gamma_{\beta}$ is given by eq.\ (\ref{effaction}).
The "dynamical factor" $\kappa$ has recently been evaluated
in terms of viscosities of the electroweak plasma \cite{carkap}.
The rate (\ref{decay})
is determined using a time independent solution of the Fokker-Planck
equation. This is justified because the characteristic time scales of
the microscopic processes (e.g., $WW$ scattering) are several orders of
magnitude smaller than the time needed for the phase transition.
Thus, the Langer theory is applicable, and it is sufficient to
use a stationary solution of the Fokker-Planck equation.

The non-trivial phase structure of the Higgs models at finite temperature
becomes only apparent if quantum fluctuations are taken into account.
The tree-level potential possesses only a single local minimum at
$\vp \neq 0$. This is analogous to the case of Coleman-Weinberg symmetry
breaking by radiative corrections. In order to obtain the
decay rate one first has to integrate out degrees of freedom which
generate the barrier between the symmetric and the broken phase and only
then, in a second step, determine the interpolating saddle point. In
\cite{bhw} it has been suggested to first integrate out the vector field.
The same route has been followed in the abelian Higgs model at zero
temperature and Coleman-Weinberg symmetry breaking \cite{eweinberg}. This
method has then been extended to the standard model \cite{bfhw} and
Yukawa models \cite{gleiser}.

Here we suggest to improve this method in the following way. In
eq.\ (\ref{decay}) the background field $\Phi$ depends only on spatial
coordinates. Therefore in the classical action $S_{\beta}$,
which appears in the integrand of the functional integral,
only the quantum field $\hat{\Phi}_0$ is shifted, i.e., the mode with zero
Matsubara frequency,
\beq \label{split}
\hat\Phi(\tau,\vec x)=\sum_n\hat\Phi_n(\vec x)
e^{\textstyle i \omega_n \tau}\quad, \quad \omega_n = 2 \pi n T \quad.
\eeq
This suggests to integrate out the scalar fields with non-zero Matsubara
frequencies in addition to the vector fields. Since the
"Compton-wave lengths" of these
modes are smaller than the wall thickness of the
critical droplet, this procedure should yield an appropriate
effective action which plays the role of the "coarse-grained"
free energy in Langer's calculation. The
scalar part of the functional integral is therefore split,
\beq \label{measure}
D\hat\Phi = \prod_n D\hat\Phi_n = D\hat\Phi_0\prod_{n\not=0}D\hat\Phi_n
\quad,
\eeq
and in the first step all modes except $\hat{\Phi}_0$ are integrated out.

The functional integral over vector fields and the non-zero
frequency modes of the scalar field yields an effective action
$\tilde{\Gamma}_{\beta}$  which replaces the classical action $S_{\beta}$
in eq.\ (\ref{decay}),
\bea \label{nonstatic}
e^{\textstyle -\Gamma_{\beta}[\Phi]}
&=&\int_\beta[D\hat{\Phi}_0]\prod_{n\not=0}\int[D\hat\Phi_n]\int_\beta[DW]
        \exp{\left(-S_\beta[\Phi+\hat{\Phi},W]
        + \int d^3x \frac{\delta \Gamma_{\beta}[\Phi]}
         {\delta \Phi(\vec{x})}\hat{\Phi}(\vec{x})\right)} \nn \\
&=&\int_\beta[D\hat{\Phi}_0]
        \exp{\left(-\tilde{\Gamma}_\beta[\Phi+\hat{\Phi}_0]
        + \int d^3x \frac{\delta \Gamma_{\beta}[\Phi]}
         {\delta \Phi(\vec{x})}\hat{\Phi}_0(\vec{x})\right)} \quad.
\eea
At one-loop order one obtains for
the effective action $\tilde{\Gamma}_{\beta}$ a sum of terms of the type
calculated in sect.\ \ref{finitet}, where now the zero-frequency
contribution of the scalar fields are absent in the loop integrals.
In addition there is a non-local contribution which stems from the
integration over the vector field,
\bea \label{grained}
\tilde{\Gamma}_{\beta}[\Phi] &=& \tilde{\Gamma}_{\beta}^{(1)}[\Phi]
+ \tilde{\Gamma}_{\beta}^{(2)}[\Phi] \quad, \nn\\
\tilde{\Gamma}_{\beta}^{(1)}[\Phi] &=& \beta \int d^3x
\left( \tilde{V}_{eff}(\vp^2,T)
+ {1\over2} (\delta_{IJ} + \tilde{Z}_{IJ}(\vp^2,T))
\vec{\nabla}\vp_I\vec{\nabla}\vp_J + \ldots\right)\quad,\nn\\
\tilde{\Gamma}_{\beta}^{(2)}[\Phi] &=& {g^2\over 2} \beta \int d^3x d^3x'
J^a_{\mu}(\vec{x})D_T^{\mu\nu}(\vec{x},\vec{x}') J^a_{\nu}(\vec{x}') \quad,
\eea
where $J^a_{\mu} = {1\over2}(\Phi^{\dagger}\partial_{\mu}\tau^a\Phi -
\partial_{\mu}\Phi^{\dagger}\tau^a\Phi)$
is the SU(2)-current and $D_{T\mu\nu}$
is the propagator for the transverse part of the vector field in the
background field $\Phi(\vec{x})$.

The difference between the effective potentials is given by
\bea \label{cubic}
\frac{\partial}{\partial\vp}\left(V_{eff}-\tilde{V}_{eff}\right)
&=&-3 T \lambda\vp \int \frac{d^3k}{(2\pi)^3} \left({1\over
\vec{k}^2+m_{\vp}^2}
+ {1\over \vec{k}^2+m_{\chi}^2}\right) \nn \\
&=&\frac{3\lambda}{4\pi}\vp T (m_{\vp}+m_{\chi} + C)\quad.
\eea
Here $C$ is a linearly divergent constant which vanishes in dimensional
regularization. We have explicitly checked that
for other regularizations it cancels against the sum of
divergent terms which arise from integration over fields with non-zero
Matsubara frequencies.
To leading order in the couplings the masses $m_{\vp}$ and $m_{\chi}$ are
given by eqs.\ (\ref{mvp}) and (\ref{mchi}).
Integrating eq.\ (\ref{cubic}) gives precisely the cubic scalar terms in
the potential (\ref{thepotential}). Hence, defining the split (\ref{split})
of the functional integral in dimensional regularization one obtains
to order $\cal{O}(g^3,\lambda^{3/2}$)
\bea \label{prepotential}
\tilde{V}_{eff}(\vp^2,T)&=&{1 \over 2}
\left({3g^2\over 16}+{\lambda\over 2}+{1 \over 4}f_t^2\right)
(T^2-T_b^2)\vp^2+{\lambda\over 4}\vp^4  \nn \\
        & &-(3 m_L^3+6 m_T^3)
\frac{T}{12\pi}+{\cal O}(g^4,\lambda^2) T_b^4 \quad,
\eea
where
\bea
m_L^2&=&{11\over 6}g^2 T^2+{g^2\vp^2 \over 4} \quad, \nn \\
m_T^2&=&\gamma^2 {g^4\over 9\pi^2} T^2 + {g^2\over 4}\vp^2 \quad.\nn
\eea
The effective potential $\tilde{V}_{eff}$ is similar
to the potential used by Dine et al.\ \cite{dine} to describe the
electroweak phase transition. For $\vp \ll T$ the term linear in T
essentially becomes the one-loop vector boson contribution reduced
by a factor $1/3$ \cite{dine}. To estimate the size of non-perturbative
corrections we have kept a non-zero magnetic mass with an order of
magnitude given by the one-loop gap equations \cite{bfhw},\cite{zwirner}.
A novel feature of the potential (\ref{prepotential})
is the way in which contributions from scalar loops are
treated. There are no terms cubic in the scalar masses, since the
integration over $\hat{\Phi}_0$ has not been carried out. However,
contrary to the treatment in \cite{dine}, the scalar-loop contributions
to the term quadratic in $\vp$ are kept, since they
arise from integration over $\hat{\Phi}_n,\ n\neq 0$.
This has an important effect on the difference between
critical temperature and barrier temperature of the transition.
For $m_H = 60$ GeV we find an increase by a factor of two.

We conclude that keeping the contribution of
scalar loops to the $T^2$-term in the effective potential and dropping
the cubic scalar mass terms yields the correct effective potential for
the zero-Matsubara frequency part of the scalar field.

As discussed in sect.\ \ref{finitet}, the part of the $Z$-factors
arises from the zero-frequency contributions to the loop integrals.
Hence, pure vector boson loops give the dominant contribution to
$\tilde{Z}$, since the integration over $\hat{\Phi}_0$ is not carried out.
The result is easily read off from
eqs.\ (\ref{zvp}),(\ref{zchi}),
\bea  \label{prez}
\tilde{Z}_{\vp}(\vp^2,T) &=& {T\over 32\pi} g^2 m^2
\left({1\over 2 m_L^3} + {5\over m_T^3}\right) \quad, \nn\\
\quad \tilde{Z}_{\chi}(\vp^2,T) &=& 0 \quad.
\eea
The "coarse-grained" effective action $\tilde{\Gamma}_{\beta}$ is now given
by eqs.\ (\ref{grained}),(\ref{prepotential}) and (\ref{prez}).
Note, $\tilde{Z}_{\vp}$ is positive. Hence, the surface tension and
therefore the strength of the phase transition is increased.
$\tilde{Z}_{\vp}$ is shown in fig.\ (\ref{wavefunction}) for three different
values of the magnetic mass.

Using eqs.\ (\ref{decay}) - (\ref{prez}) we can now calculate
the decay rate of the metastable symmetric phase.
As discussed in sect.\ 2, the $Z$-factors $Z_{\vp,\chi}$ are expected to yield
only small corrections. Therefore we start from $Z=0$ and the effective
potential (\ref{prepotential}) and determine a saddle point $\bar{\vp}$
which interpolates between the symmetric and the broken phase. The
remaining $\hat{\Phi}_0$-integration is then carried out in the Gaussian
approximation. The spectrum of fluctuations contains one negative
eigenvalue $\lambda_-$ which guarantees that $Z_{\beta}[\bar{\vp}]$
is purely imaginary. Further, there are
six zero modes from translational invariance and global SU(2)$\times$U(1)
symmetry, and a discrete and continuous positive spectrum. The contribution
of the zero-modes to the fluctuation determinant is given by \cite{bfhw}
\bea \label{volume}
{\cal V}_{trans} V&=&\left({1 \over 2\pi}
\tilde \Gamma_\beta[\bar{\Phi}]\right)^{3/2}V,\nn\\
{\cal V}_{int}&=&\frac{\pi^2}{2}
\left({\beta\over 2\pi}\int d^3x\,\bar{\vp}^2\right)^{3/2},
\eea
where $V$ is the total volume of the physical three-dimensional space.
In the thin wall approximation also the contribution from
``Goldstone modes''  has been evaluated \cite{bfhw}, which correspond to
deformations of the droplet surface. However,
for the electroweak phase transition the thin wall approximation is
known not to be reliable \cite{dine},\cite{carkap}. We therefore
base our calculations on the decay rate (cf.\ (\ref{decay}), (\ref{volume}))
\beq \label{rate}
\frac{\Gamma}{V}= \kappa \mid\lambda_-\mid^{-1/2} \
\cal{V}_{trans}\ \cal{V}_{int}\ \mu^7\
e^{\textstyle -\tilde{\Gamma}_{\beta}[\bar{\Phi}]}
= \ A\  e^{\textstyle -\tilde{\Gamma}_{\beta}[\bar{\Phi}]} \quad,
\eeq
where $\bar{\vp}$ is the saddle point and $\mu = m_{\vp}(0,T)$.
The negative eigenvalue $\lambda_-$ is approximated by $-2/R^2$, its
thin wall value, where the radius $R$ of the critical droplet is
defined by $\partial^2\bar{\vp}(r=R,T)/\partial r^2  = 0$.
$\kappa$ can be estimated using the results given in \cite{carkap}.

We have numerically determined the saddle point $\bar{\vp}$
at temperatures close to the critical temperature $T_c$ for different
values of the scalar self-coupling $\lambda$ and the parameter $\gamma$,
i.e., the magnetic mass $m_T$. This will be described in more detail in
the following section. Fig.\ (\ref{prefactor})
shows the ratio of the pre-factor in
eq.\ (\ref{rate}) normalized to $T^4$ as function of the Higgs mass
at the respective temperature $T_e$ where
the phase transition is completed. Clearly, the logarithm of the
pre-factor is much smaller than $\Gamma_{\beta}[\bar{\vp}]$ which,
at the temperature $T_e$, is $\sim 140$. Hence, the semiclassical
approximation is justified. The difference between the pre-factor $A$
and $T^4$ is numerically unimportant for the description of the
cosmological phase transition. As the Higgs mass $m_H$ approaches the
critical mass $m_H^{crit}$, where the transition becomes second-order
for $\gamma \neq 0$, the pre-factor strongly decreases. This is mostly
due to the decrease of $\mu$.

A measure for the size of the one-loop correction to the $Z$-factor
is the ratio
\beq
\delta_Z = \frac{\int d^3x \tilde{Z}_{\bar{\vp}}
            \left(\vec{\nabla}\bar{\vp}\right)^2}
           {\int d^3x \left(\vec{\nabla}\bar{\vp}\right)^2} \quad.
\eeq
This correction is shown in fig.\ (\ref{zcorrection}) as function of
the Higgs mass for different
values of $\gamma$. For $\gamma=0$ the perturbative expansion breaks down
at $m_H \sim 80$ GeV, in agreement with results obtained previously
\cite{bfhw}. For $\gamma=1,2$ the correction never exceeds 10\% which
demonstrates the importance of the infrared cutoff provided by the
magnetic mass.

The effective potential to order $\cal{O}(g^3,\lambda^{3/2})$ is given
by eq.\ (\ref{thepotential}). The corresponding quantity without the
one-loop contribution of the static part of the scalar field $\Phi$
is $\tilde{V}_{eff}$ (\ref{prepotential}). A more accurate approximation
to the effective potential can be obtained by performing the
$\hat{\Phi}_0$-integration over the expression (\ref{nonstatic}) in
the Gaussian approximation. The result is the potential
\beq \label{imapotential}
\bar{V}_{eff} = \tilde{V}_{eff} - \frac{T}{12\pi}(m_{\vp}^3+3 m_{\chi}^3)
\quad,
\eeq
where, contrary to eq. (\ref{thepotential}), the scalar masses are now
given by
\beq \label{imamasses}
m^2_{\vp}=\frac{\partial^2 \tilde{V}_{eff}(\vp^2)}{\partial \vp^2}\,  ,\quad
m^2_{\chi}={1\over\vp}\frac{\partial \tilde{V}_{eff}(\vp^2)}{\partial \vp}
\quad.
\eeq
These masses agree with the  mass terms (\ref{mvp}) and (\ref{mchi})
only to leading order in the couplings. The potentials $V_{eff}$ and
$\bar{V}_{eff}$ clearly differ by an infinite number of terms, if
the masses (\ref{imamasses}) are formally expanded in powers of $\lambda$.
The most important property of the mass terms (\ref{imamasses}) is that they
become negative as the scalar field varies between the two local
minima of $\tilde{V}_{eff}$. Hence, the true effective potential
$\bar{V}$ is complex. This general feature of non-convex potentials,
which is well known for models at zero temperature, is not visible to
any finite order in perturbation theory.

What is the meaning of the imaginary part of the
effective potential? At zero temperature the imaginary part is defined
by first substituting
\beq
m^2(\vp) \rightarrow m^2(\vp) - i\epsilon \quad,
\eeq
and then performing the limit $\epsilon \rightarrow 0$. This yields the
imaginary part
\bea
\cal{I}(\vp) &=& 2\ \mbox{Im} V(\vp) \\
             &=& {1\over 32\pi} \mid m^2(\vp)\mid^2 \Theta(-m^2(\vp))
\eea
which is interpreted as decay rate per unit time and volume
of the homogeneous state into inhomogeneous states of
lower energy \cite{weinwu}.

It is straightforward to adopt the same procedure at finite temperature.
One immediately obtains from the cubic scalar mass terms
in eq.\ (\ref{imapotential})
\bea \label{homt}
\cal{I}_{\beta}(\vp) &=& 2\ Im \bar{V}_{eff}(\vp,T) \\
                     &=& {T\over 6\pi}\
( \mid m_{\vp}^2\mid^{3/2}\Theta(-m_{\vp}^2)
  +\mid m_{\chi}^2\mid^{3/2}\Theta(-m_{\chi}^2)) \quad.
\eea
This should then be the finite-temperature decay width of a homogeneous state
into mixed states. This interpretation appears sensible at a stationary
point $\vp$ of the potential $\bar{V}(\vp^2,T)$
where $m_{\chi}=0$, i.e., eq.\ (\ref{homt})
should be applicable at the top of the
barrier. For arbitrary values of $\vp$ the interpretation
of the rate (\ref{homt}) is less clear due to the gauge dependence
of this expression.

\section{The cosmological phase transition} \label{cosmopt}

We can now use the results obtained in the last section to investigate the
cosmological electroweak phase transition. This phase transition
has already been studied in considerable detail by several groups
\cite{dine},\cite{bfhw},\cite{enqvist}-\cite{carkap}.
Here we will extend the previous studies mainly in two respects.
We shall study the dependence of the transition on the mass of the Higgs
boson, and we shall also investigate the influence of a non-zero
magnetic mass. In this case the transition becomes second order for
Higgs masses above a critical mass $m_H^{crit}$ \cite{bfhw}, and the reheating
of the symmetric phase can qualitatively change the nature of the transition.

The main features of the transition can be understood by
considering the simplified potential
\beq \label{simplepot}
V=\frac 1 2 a (T^2-T_b^2) \vp^2 - \frac 1 3 b T \vp^3 +
\frac 1 4 \lambda \vp^4 \quad,
\eeq
where the parameters $a$ and $b$ can be read off from
eq.\ (\ref{prepotential}). Here we neglect $m_L$, which contributes
essentially only to the $T^4$-term, and set $\gamma=0$.
An analytical treatment of the transition is possible in the thin wall
approximation \cite{linde}. Quantitatively, this approximation is too
crude for the electroweak phase transition. However, it
is still useful to get a qualitative picture of the transition and to obtain
the order of magnitude of the various parameters involved. This
will be confirmed by our numerical calculations discussed below.

In the thin wall approximation the nucleation rate is given by
\beq
\frac{\Gamma(t)}{V}=A \exp{\left(-B \left(\frac {t_c}{t_c-t}\right)^2
\right)}\quad,
\eeq
where
\beq
A=\omega T_b^4 \quad,\quad
B=\frac{\sqrt{2}\  2^6\ \pi}{3^{20}}\frac{b^5}{a^2 \lambda^{7/2}}\quad,
\eeq
with $\omega = \cal{O}(1)$. $T_b$ is the barrier temperature, and $t_c$
is the time at which the critical temperature $T_c \approx T_b$ is reached.
Using eqs.\ (\ref{Tb}), (\ref{prepotential}) and the relation between time
and temperature, $t \approx 0.03\  m_{PL}/T^2$ with $m_{PL}=1.2\cdot
10^{19}$ GeV \cite{kolbturn}, one obtains the relations
\bea
T_b &=& 120\ \mbox{GeV}\ G_F^{-1/2}\ (3 m_W^2 + m_H^2 +2 m_t^2)^{-1/2}
\left({m_H\over 100 \mbox{GeV}}\right) \quad,\\
\nn  \\
t_c &=& 1.7\cdot 10^{-11}\ \mbox{sec}\ G_F \ (3 m_W^2 + m_H^2 + 2 m_t^2)
\left({100 \mbox{GeV} \over m_H}\right)^2 \quad,\label{tc}\\
\nn \\
B &=& \frac{2^{10}}{3^4\ \pi^4}
\frac{G_F^2\ m_W^{15}}{(3 m_W^2 + m_H^2 + 2 m_t^2)^2\ m_H^7} \quad.\label{b}\\
\eea
{}From these equations one can read off
the involved orders of magnitude of time and temperature
and also the dependence on the Higgs boson mass.

As the universe expands the temperature $T$ decreases. Once the critical
temperature $T_c$ is reached critical droplets can start to nucleate
and grow. At time $t$ the number density of droplets of size $r$ is
\beq
n(t,r)=\Gamma\left(t-{r\over v}\right) \quad.
\eeq
Here $v$ is the velocity of the droplet wall, and the initial size of
the critical droplet has been neglected.
The fraction of space filled with the new phase is then
\beq \label{fraction}
f_B(t)=\exp{\left(-{4\pi \over 3}
\int_0^{r_{max}(t)}dr r^3 n(t,r)\right)} \quad,
\quad r_{max}(t) = v (t - t_c) \quad,
\eeq
where the exponentiation in eq.\ (\ref{fraction}) accounts for
the overlap between different droplets.

The phase transition is completed at time $t_e$, which one may define
implicitly by $f_B(t_e) = 1/e$. The nucleation rate is non-negligible
only for times very close to $t_e$. This allows one to derive simple
expressions for the time $t_e$, the ``transition time''  $\tau$ during which
the phase transition essentially takes place,  the total number of
droplets $N(t)$ per unit volume
and the average radius $R(t)$, which are defined as
\bea
N(t) &=& \int_0^{r_{max}(t)} dr\ n(t,r)\quad,\\
\bar{R}(t) &=& {1\over N(t)} \int_0^{r_{max}(t)} dr\ r \ n(t,r) \quad.
\eea
A straightforward calculation yields for the time $t_e$ the implicit
relation
\beq \label{te}
{\pi\over 2}\frac{A\ t_c^4}{B^4} \left(\frac{t_e-t_c}{t_c}\right)^{12}
\exp{\left(-B\left(\frac{t_c}{t_e-t_c}\right)^2\right)} = 1 \quad,\\
\eeq
and for the transition time $\tau$ one obtains
\beq \label{tau}
\tau = \frac{(t_e-t_c)^3}{2 B t_c^2} \quad.
\eeq
The fraction of converted space, the number of droplets and the
average radius are given by
\bea
f_B(t) &\approx& \exp{\left(-\exp{\left(\frac{t_e-t}{\tau}\right)}\right)}
\quad,\\
N(t) &\approx& {1\over 8\pi \tau^3}
\exp{\left(\frac{t_e-t}{\tau}\right)} \quad, \\
\bar{R}(t) &\approx& \tau \quad.
\eea
These equations demonstrate that the phase transition takes place in
the time interval $\tau$ before $t_e$. Using eqs.\ (\ref{tc}),(\ref{b}),
(\ref{te}) and (\ref{tau}) one easily verifies the
following hierarchy between the different time scales
\beq
t_c \gg t_e - t_c \gg \tau \quad.
\eeq

We have also studied the phase transition numerically. Here the free
energy of the critical droplet is calculated using the potential
$\tilde{V}_{eff}$ (cf.\ (\ref{prepotential})) derived in sect.\ \ref{falsev}.
We keep a non-zero magnetic mass of the size given by the one-loop
gap equations, i.e., we choose $\gamma = 1$ in eq. (\ref{truemt}).
For the potential (\ref{prepotential}) the stationary
point $\bar{\vp}$ and the corresponding free energy are determined
numerically for temperatures below $T_c$ down to $T_e$, where the transition
is completed. The course of the transition depends on the velocity of
the droplet wall which has been discussed in detail in the literature
\cite{dine},\cite{turok},\cite{mclerran}. Here we just
illustrate the dependence by calculating the course of the transition
for three different velocities, $v = c$, $v = c/10$ and
$v = c/1000$. The initial radius of the critical droplet, which is very
small (cf. \cite{bfhw}), is also taken into account.

Figs.\ (\ref{h30f}) - (\ref{h30T}) show the result
for $m_H = 30$ GeV, which we have chosen as an example of a small Higgs mass.
The completion time $t_e$ is $\sim 3\cdot 10^{-12}$ sec
and depends only weakly on the
wall velocity (fig.\ (\ref{h30f})). As one may have expected,
the number of droplets (fig.\ (\ref{h30N}))
and the average radius (fig.\ (\ref{h30R})) at the end of the transition scale
approximately like $v^{-3}$ and $v^{-1}$.
The number density of droplets is given per cm$^3$. Note, that the size of
the horizon at the critical temperature is of order 1 cm.
Following Enqvist et al.\ \cite{enqvist}
and Carrington and Kapusta \cite{carkap} we have plotted in fig.\ (\ref{h30T})
the change in temperature as function of time. The system supercools
from time $t_c$ until the time $t_s$ at which the supercooling
stops and the liberated latent heat leads
to some reheating. The increase in temperature can be estimated by
\beq
\Delta T_R \ =\ {1\over c_v}\ L \quad,
\eeq
where the latent heat is given by the derivative of the effective potential,
\beq
L\ = \ \left.
T {\partial\over \partial T} \tilde{V}_{eff}(T,\vp^2)\right|_{T=T_c} \quad,
\eeq
and the specific heat of the symmetric phase is
\beq
c_v\ =\ {d\over dT} e(T) \quad, \quad e(T)\ =\ {427\over 120} \pi^2 T^4 \quad.
\eeq
Here $e(T)$ is the energy density of the standard model at high temperatures
to leading order in the coupling constants,
where gluons and the U(1)-gauge boson have been included. The above equations
are easily derived from the usual thermodynamic relations and
the relations for the pressure
$p_s(T) = - \tilde{V}_{eff}(T,0)$ and
$p_b(T) = - \tilde{V}_{eff}(T,\vp^2(T))$
in the symmetric and broken phase, respectively.
Using these equations to compute the reheating temperature assumes that the
wall velocity is sufficiently small so that the latent heat can indeed be
converted into thermal energy of the symmetric phase.

In fig.\ (\ref{h30T}) the true evolution is compared with a transition where
thermal equilibrium is maintained at all times. In this
hypothetical case the
temperature remains at $T_c$ until the time $t_e'$, at which all latent
heat has been used up to expand the universe. At later times the cooling due
to the expansion then proceeds in the usual manner. Note that
the real phase transition starts after considerable supercooling
and completes in a very short time. The temperature
after reheating is slightly larger than the temperature of the adiabatic
transition at the same time. This is due to the entropy generated in the
transition.

For the Higgs mass $m_H = 60$ GeV the transition looks qualitatively
the same. The time interval $t_s-t_c$
decreases by about two orders
of magnitude to $\sim 2\cdot 10^{-14}$ sec (fig.\ (\ref{h60})).
Correspondingly, the average radius decreases by about two orders
of magnitude whereas the number density of droplets increases
by about six orders.
Since much less time is available for supercooling, the temperature
after reheating is now closer to the critical temperature $T_c$ than
for $m_H = 30$ GeV. Our numerical results for $m_H = 60$ GeV
are consistent with those obtained in
\cite{carkap}, given the differences in the choice of $m_t$,
the magnetic mass and the wall velocity.

A dramatic change occurs for $m_H = 80$ GeV (fig.\ (\ref{h80})).
After the initial supercooling period the liberated
latent heat starts to reheat the system
at the time $t_s < t_e'$, long
before the transition is completed. Due to the small supercooling
the latent heat can reheat the system to
the critical temperature.  After that no new droplets
nucleate, the average radius $\bar{R}$ increases linear
with time, and the transition completes in equilibrium
(fig.\ (\ref{h80})), resembling
the QCD phase transition \cite{enqvist}. Compared to $m_H = 60$ GeV
the time interval $t_s-t_c$ has decreased again by two orders of magnitude
to $\sim 3\cdot 10^{-16}$ sec.
Contrary to the QCD phase transition the time difference $t_e'-t_c$ is
much smaller than the Hubble time due to the smallness of the latent heat.

The change of the nature of the phase transition for large values of the Higgs
mass is a consequence of the magnetic mass.
For any finite value of this mass, the phase transition becomes
very weakly first-order with rising Higgs mass and
eventually second-order above a
critical Higgs mass $m_H^{crit}$ \cite{bfhw}. Our calculations show that below
this critical
value there is a large range of Higgs masses for which the universe reheats
up to $T_c$. This is shown in fig.\ (\ref{reheat})
where the reheating temperature $\Delta T_R$
is compared with the supercooling temperature needed to initiate the
phase transition
$\Delta T_{SC}$. Clearly,
the reheating temperature reaches $T_c$ if $\Delta T_R > \Delta T_{SC}$.

This interesting result can also be understood analytically. Consider
a potential with a generic plasma mass, in our case the
simplified potential (\ref{simplepot}) with non-zero magnetic mass,
\begin{equation}\label{simplemt}
V=\frac 1 2 a(T^2-T_b^2)\vp^2- \frac 1 3 b T (c^2 T^2 +\vp^2)^{3/2}
+\frac 1 4 \lambda \vp^4 \quad.
\end{equation}
This potential implies a change from a first-order to a second-order
phase transition as the Higgs mass increases.
Since there are strong hints that the magnetic mass is non-zero, we
expect that eq.\ (\ref{simplemt})
is a reasonable approximation
for the effective potential of the standard model.
Above a critical value of the scalar coupling,
\begin{equation}
\lambda_{crit}=\frac{b} {2 c}\quad,
\end{equation}
the potential  (\ref{simplemt}) describes a second-order phase transition. For
values of
$\lambda$ slightly below $\lambda_{crit}$ the transition is weakly first-order.
We are interested in the behaviour of several quantities as $\lambda$
approaches $\lambda_{crit}$ from below. The position $\vp_c$ of the degenerate
minimum at $T_c$ vanishes in this limit. Some other characteristic quantities
of the first-order transition vanish as powers of $\vp_c$. We find for the mass
difference $\Delta m$, the
maximum of the effective potential ${V}^{max}$,
the surface tension $\sigma$, the transition rate $\Gamma$, the latent heat $L$
and
the temperature differences $T_c-T_b$ and $T_c-T_s$, where $T_s$ is the
temperature corresponding to $t_s$,
\begin{eqnarray}
\Delta m\equiv m_H^{crit}-m_H&\sim& \vp_c^2\quad ,\nn \\
{V}^{max}&\sim&\vp_c^6 \quad,\nn \\
\sigma&\sim& \vp_c^4 \quad , \nn \\
 L&\sim&\vp_c^2 \quad,\nn \\
\Gamma &\sim& \vp^0 \quad,\nn \\
T_c-T_b&\sim&\vp_c^2 \quad , \nn \\
T_c-T_s&\sim&\vp_c^4 \quad.
\end{eqnarray}
Using these equations one obtains for  $t_s$,
the time at which the supercooling stops for the true transition, and for
$t_e'$, the time for which the equilibrium transitions complete, respectively,
\begin{eqnarray}
t_s-t_c&\sim&T_c-T_s\sim\vp_c^4\sim \Delta m^2 \quad,\nn \\
t_e'-t_c&\sim&L\sim\vp_c^2\sim \Delta m \quad.
\end{eqnarray}
Clearly, for Higgs masses very close to $m_H^{crit}$ one has
$t_s<t_e'$, i.e., the supercooling ends,
the liberation of the latent heat reheats the system to $T_c$ and the phase
transition completes in equilibrium. For vanishing magnetic mass
this phenomenon does not occur for Higgs masses smaller than the
vector boson mass, for which the perturbative
approach is reliable.

\section{Summary}
In this paper we have extended previous studies of the electroweak phase
transition in several respects. In particular, we have studied in
greater detail the decay rate for the metastable symmetric phase
and also the course of the phase transition.
On the whole the by now familiar picture
is confirmed that the phase transition is
weakly first-order for Higgs boson masses up to the W-boson mass
where the perturbative approach becomes unreliable.

We have calculated the two wave function correction factors $Z_{\vp}$
and $Z_{\chi}$ for the SU(2)-Higgs model, and we have evaluated the
corresponding correction to the free energy of critical droplets.
We find that the perturbative expansion
is reliable for Higgs masses below $80$ GeV.

The evaluation of the decay rate of the metastable symmetric phase is
a non-trivial problem, especially since the metastability is entirely
due to quantum corrections. Hence, two expansions are needed, first
an improved perturbative expansion to obtain the ``coarse-grained''
effective action with a barrier between the two local minima,
and second the expansion around the critical droplet, the saddle
point of the ``coarse-grained'' effective action. We have improved
previous calculations by integrating out the vector fields and the
components of the scalar fields with non-zero Matsubara frequencies
in the first step. However, further work is still needed
to obtain a better understanding of
renormalization, gauge dependence and convergence of the performed
expansions.

As a by-product of our calculation we have derived a formula for the
decay rate of a homogeneous unstable state, which is the finite-temperature
analogue of the familiar zero-temperature decay rate. Both formulae
are obtained from the imaginary part of the effective potential. The
interpretation of the formula away from a local maximum of the potential,
in particular its gauge dependence, requires further study.

Our numerical analysis of the course of the cosmological electroweak
phase transition shows that at a Higgs boson mass of $\sim 60$ GeV
a qualitative change of the nature of the transition occurs. This is
a consequence of the very weak first-order phase transition which
in our case is due to the assumed non-zero magnetic mass of the W-boson.
For Higgs masses
above 60 GeV the reheating temperature reaches the critical
temperature before the transition is completed. The phase transition
then continues in equilibrium, similar to the QCD phase transition.
This may have interesting implications for the generated density fluctuations.

As the Higgs mass increases the phase transition becomes more and more
dependent on non-perturbative properties of the symmetric phase. In our
calculations this is reflected in the strong dependence on the value
of the magnetic mass of the vector boson. Recent lattice calculations
indicate that such non-perturbative effects may increase the strength
of the first-order transition. A deeper understanding of these
non-perturbative effects appears to be a crucial step on the way towards
a theory of the electroweak phase transition.

The work of D. B. has been supported by the ``Graduiertenkolleg
f\"ur theoretische Elementarteilchenphysik'', Universit\"at Hamburg.
Z. F. acknowledges partial support from Hung. Sci. Grant under Contract
No. OTKA-F1041.

\vfill\eject

\appendix{\bf Appendix A \, Contributions to $Z(\Phi,T))$}
\renewcommand{\theequation}{\mbox{A}.\arabic{equation}}
\setcounter{equation}{0}

As it was promised we show that only graphs of the type shown in fig.\ 1
with plasma mass insertions on the internal lines give contributions
in leading order to $Z(\Phi,T)$.

Instead of $g$, $\sqrt\lambda$ and $f_t$ we use in this appendix a generic
coupling $h$.

A diagram can contribute to an order in $h$ lower than the power
in $h$ obtained from vertices,
if some of the Matsubara
frequencies of the loop variables vanish.
In the following we call these loops
and the corresponding lines {\underline {soft}}.
Consider a contribution to the effective
potential from a graph $G$ of the following type:
\hfill\break
a. it has at least one internal scalar line $l_{scalar}$;\hfill\break
b. by cutting $l_{scalar}$ the diagram remains
one-particle irreducible; \hfill\break
c. $r$ loop variables (among them the momentum of $l_{scalar}$) are
soft; \hfill\break
d. the loops of non-zero Matsubara frequencies
(in the following {\underline {hard}} loops)
yield leading order plasma mass corrections (cf. fig.\ 12, where thick
lines are hard, thin ones are soft, $r=2$).

According to appendix A of \cite{bfhw}
this graph gives a contribution ${V}_G$
to the effective potential of order $h^{r+2}$.
Writing ${V}_G$ as a Feynman integral
over $\vec {p}$ (the momentum of $l_{scalar}$) with the
appropriate self-energy insertion
$\Sigma_G(p_0=0,{\vec{p}}^2)$
and performing a Taylor-expansion in ${\vec p}^2$
one gets
\begin{eqnarray}
{V}_G&=&\int {d^3p \over (2\pi)^3}
{1 \over {\vec{p}}^2+m^2} \Sigma_G(p_0=0,{\vec{p}}^2)
\nonumber\\
&=&\int {d^3 p \over (2\pi)^3} \frac 1 {{\vec{p}}^2+m^2}
\left[\Sigma_G(p_0=0,{\vec{p}}^2=0)
+{\vec{p}}^2\Sigma_G'(p_0=0,{\vec{p}}^2=0)+\ .\ .\ .\  \right]
\nonumber\\
&=&C+h^3\int{d^3 x \over (2\pi)^3}
\frac 1 {{\vec{x}}^2+1}\vec{x}^2\Sigma'_G(p_0=0,{\vec{p}}^2=0),
\end{eqnarray}
where $\Sigma_G'=\partial \Sigma_G/\partial {\vec p}^2$,
and we have used the substitution $\vec{p}=m \vec{x}$
(remember that $m$ is of order $h$).
Since ${V}_G$ is of order $h^{r+2}$ and the
above integral over $\vec x$ is independent of $h$,
the derivative term
$\Sigma'(p_0=0,\vec{p}^2=0)$ is of order $h^{r-1}$.
In our case one-soft-loop contributions
with plasma mass corrections to
$\Sigma(p_0=0,{\vec{p}}^2)$ correspond to
two-soft-loop ($r=2$) contributions to ${V}_G$,
thus the wave function correction term
is at least of order $h^{2-1}=h$.
Summing all plasma mass corrections yields the full propagators
of the improved perturbation theory for internal lines.
This $\cal{O}(h)$
contribution is non-zero and it is
given by eqs. (\ref{zvp}),(\ref{zchi}).

According to (A.16) of \cite{bfhw} other diagrams
(eg., diagrams with more soft loops
or more complicated hard loop structures
then  mentioned above) lead to higher order corrections to
$Z(\Phi,T)$.

For completeness we
also list the $Z$-factors for a model with $n$ complex fields
$\Phi=(\Phi^{(1)},...,\Phi^{(n)})$,
$\Phi^{(\alpha)}=(\varphi^{(\alpha)}_1+i\varphi^{(\alpha)}_2)\sqrt{2}$,
where the diagonal global $U(1)$ symmetry is gauged. At one loop
order, zero-Matsubara frequency internal lines yield the wave function
correction terms (cf.\ fig.\ 1.\ a,b):
\bea
&&{1\over2}Z_{IJ}^{(\alpha \beta)}(\Phi,T)
\vec{\nabla}\vp^{(\alpha)}_I\vec{\nabla}\vp^{(\beta)}_J = \\
&-&{1\over4} (\Phi^{\dagger}\Phi)^{-1}
(\vec{\nabla}\Phi^{\dagger}\Phi-\Phi^{\dagger}\vec{\nabla}\Phi)^2Z_1
+\vec{\nabla}\Phi^{\dagger}\vec{\nabla}\Phi Z_2
+{1\over4} (\Phi^{\dagger}\Phi)^{-1}
(\vec{\nabla}(\Phi^{\dagger}\Phi))^2 Z_3,\nn
\eea
where
\bea
Z_1(\varphi^2,T)&=&{2e^2T \over 3\pi}
\left( {1 \over m_\varphi+m_T} - {1 \over m_\chi+m_T}\right)\nn\\
Z_2(\varphi^2,T)&=&-{2e^2T \over 3\pi}{1 \over m_\chi+m_T}\nn\\
Z_3(\varphi^2,T)&=&{e^2Tm^2 \over 48\pi}
\left( {1 \over m_L^3}+{10 \over m_T^3}\right),
\eea
$e$ is the gauge coupling and $m=e\varphi$. For $n=1$, the usual
abelian Higgs model, only two linear combinations contribute,
$Z_\chi=Z_2-Z_1$ and $Z_\varphi=Z_2+Z_3$.

\vfill\eject

\vfill \eject

\renewcommand{\thesection}{\ }
\section{Figure captions}

\begin{description}
\item[Figure \ref{zfac}:] One-loop contributions
to the wave function correction factors.

\item[Figure \ref{wavefunction}:] The wave function correction factor
$\tilde{Z}_{\vp}$ for three different values of the magnetic mass.

\item[Figure \ref{prefactor}:] The pre-factor of the decay rate as
function of the Higgs boson mass.

\item[Figure \ref{zcorrection}:] One-loop wave function correction
as function of the Higgs boson mass $m_H$ for different values of $\gamma$.

\item[Figure \ref{h30f}:] Volume fraction of the new phase as function
of $t - t_c$; $m_H = 30$ GeV.

\item[Figure \ref{h30N}:] Number density of droplets per cm$^3$ as function of
$t - t_c$; $m_H = 30$ GeV.

\item[Figure \ref{h30R}:] Average droplet radius as function of
$t - t_c$; $m_H = 30$ GeV.

\item[Figure \ref{h30T}:] Change in temperature of the true (solid line)
and the equilibrium (dashed line) transitions as function of
$t - t_c$; $m_H = 30$ GeV.

\item[Figure \ref{h60}:]  Change in temperature of the true (solid line)
and the equilibrium (dashed line) transitions as function of
$t - t_c$; $m_H = 60$ GeV.

\item[Figure \ref{h80}:]
 Change in temperature of the true (solid line)
and the equilibrium (dashed line) transitions as function of
$t - t_c$; $m_H = 80$ GeV.

\item[Figure \ref{reheat}:] Comparison of the temperature decrease due to
supercooling and the temperature increase due to reheating as
function of the Higgs boson mass.

\item[Figure \ref{graph}:] A two-soft-loop contribution to the
effectiv potential.
\end{description}
\nopagebreak
\begin{list}{}{\usecounter{figure}}
\item \label{zfac}
\item \label{wavefunction}
\item \label{prefactor}
\item \label{zcorrection}
\item \label{h30f}
\item \label{h30N}
\item \label{h30R}
\item \label{h30T}
\item \label{h60}
\item \label{h80}
\item \label{reheat}
\item \label{graph}
\end{list}
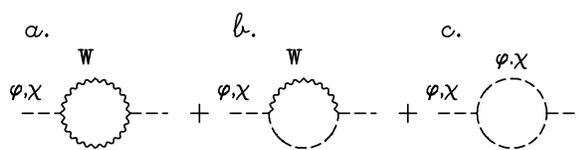
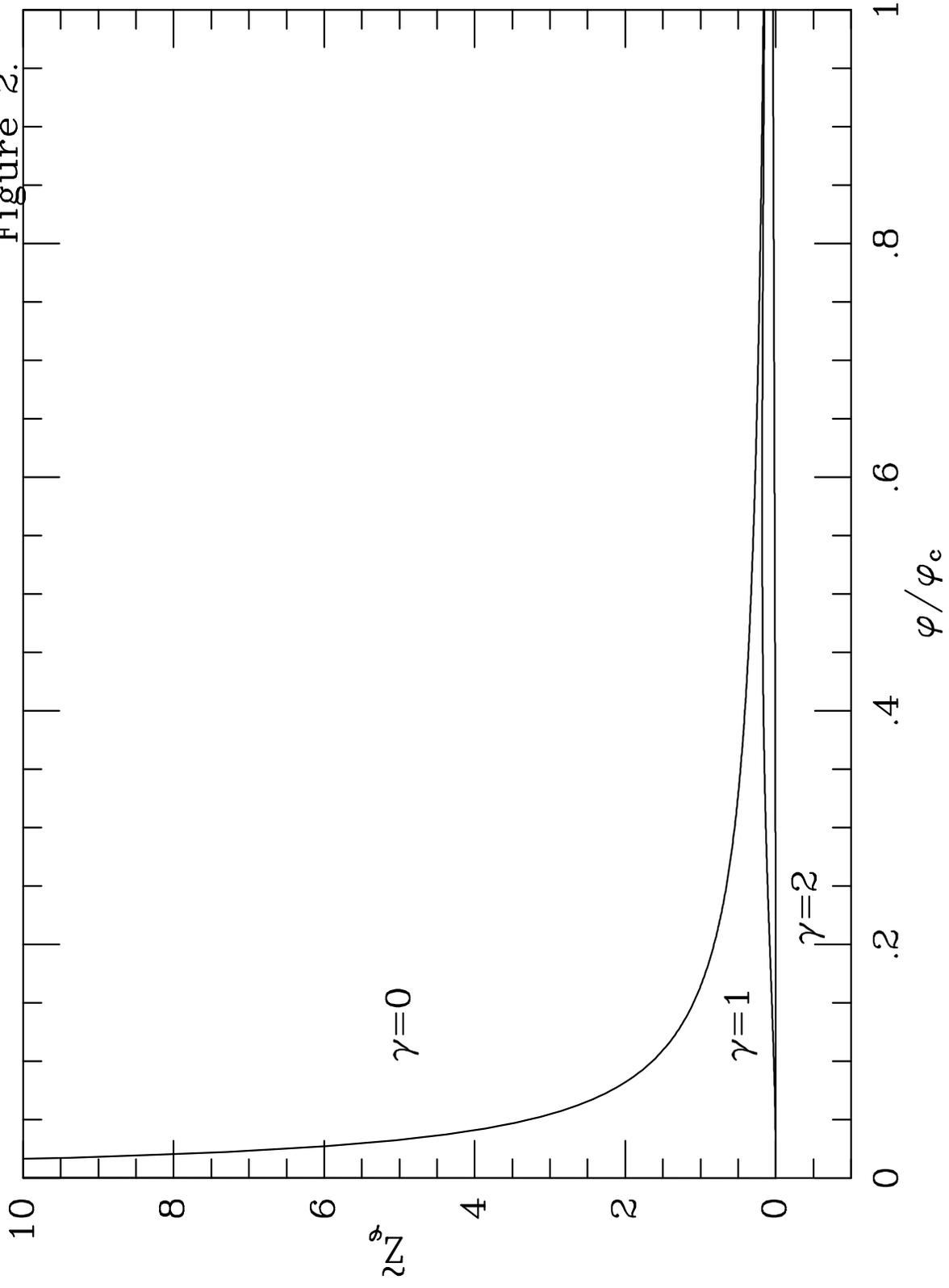
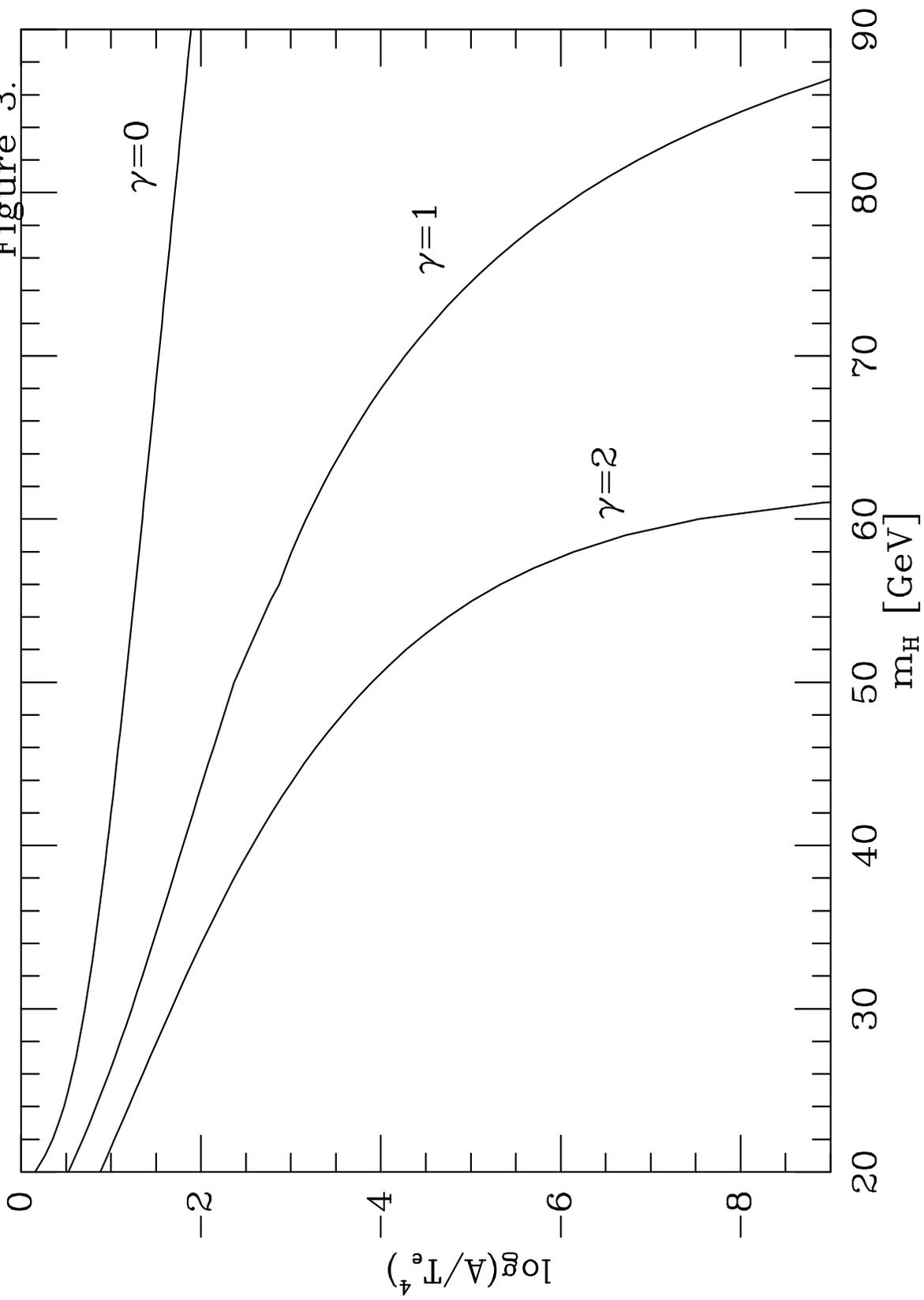
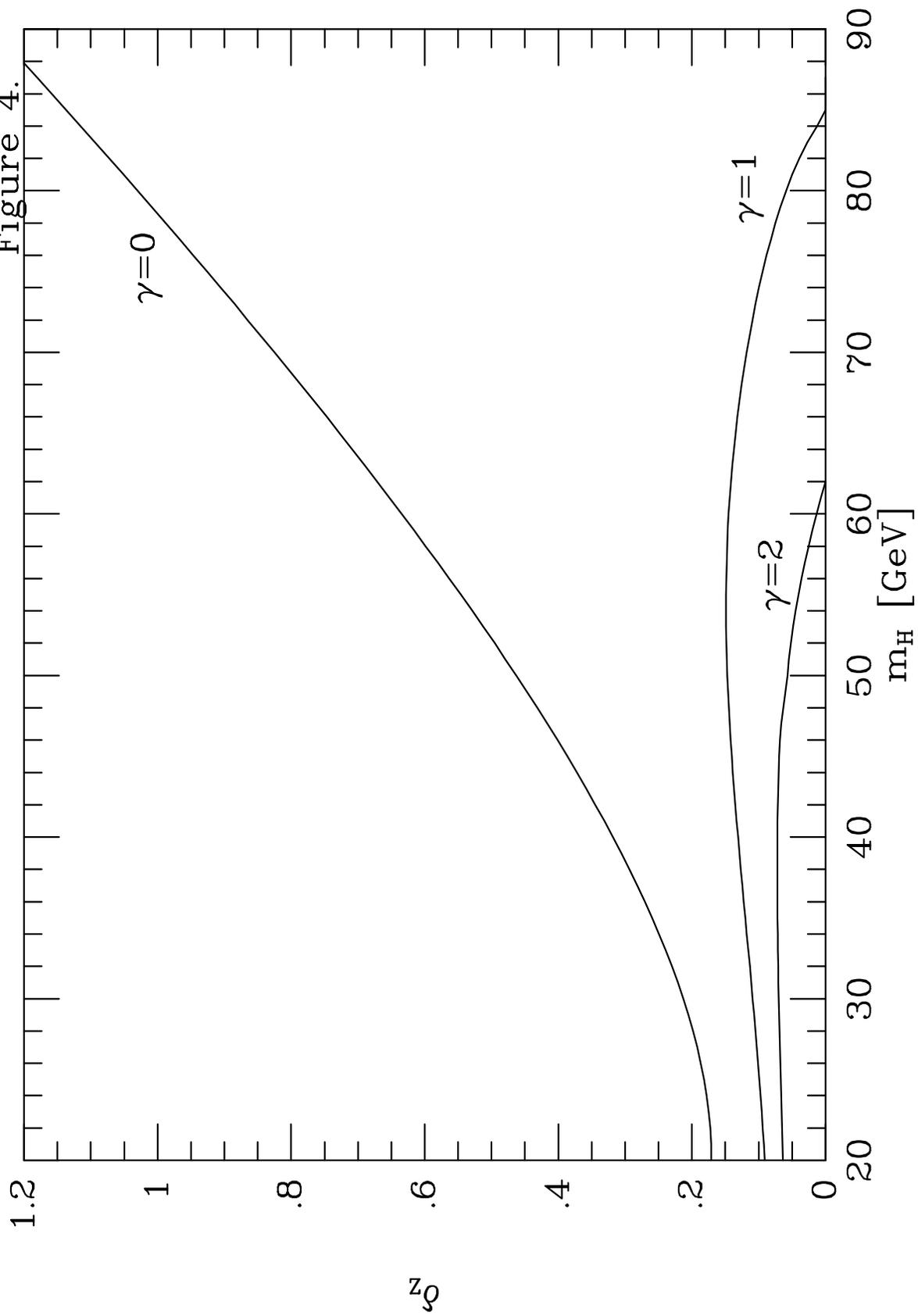
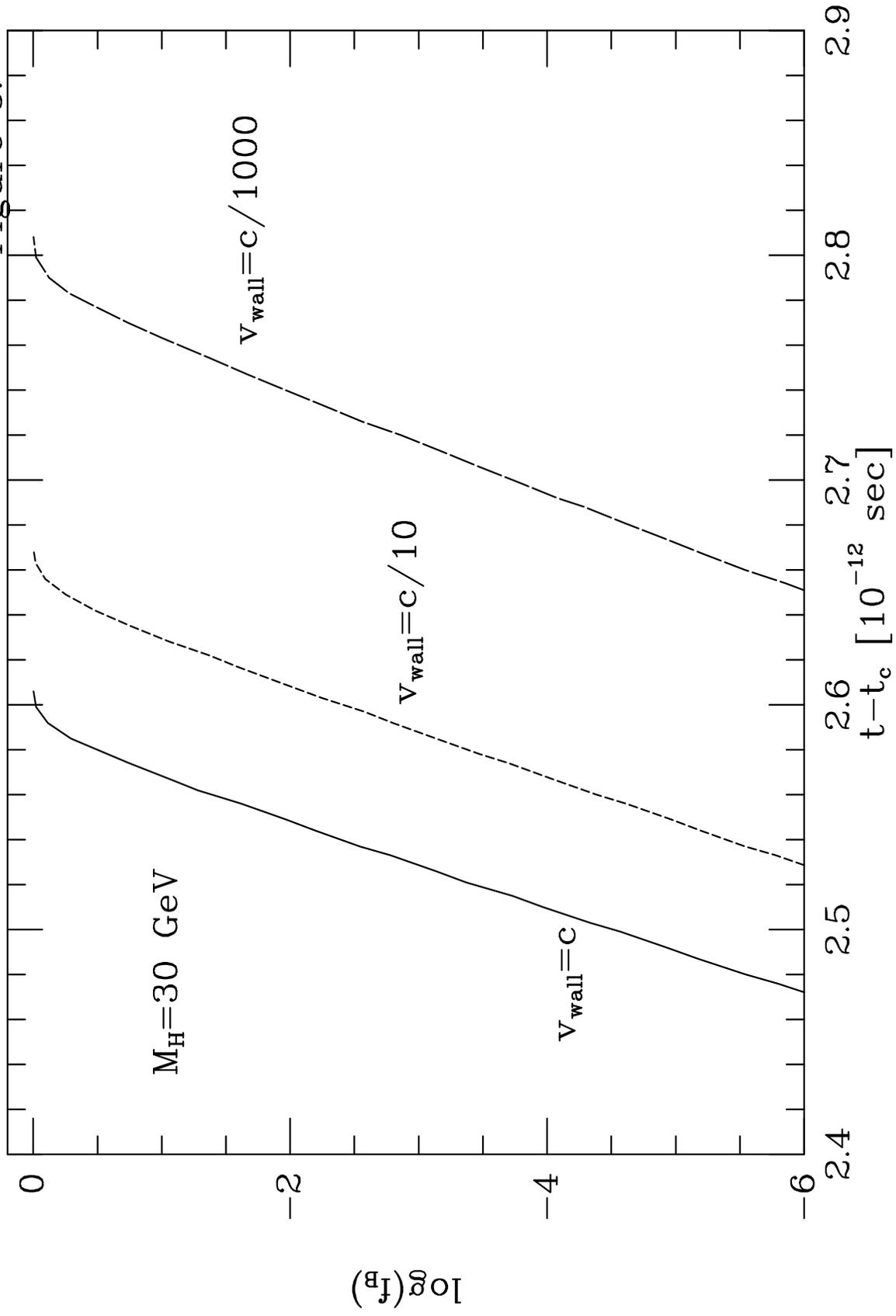
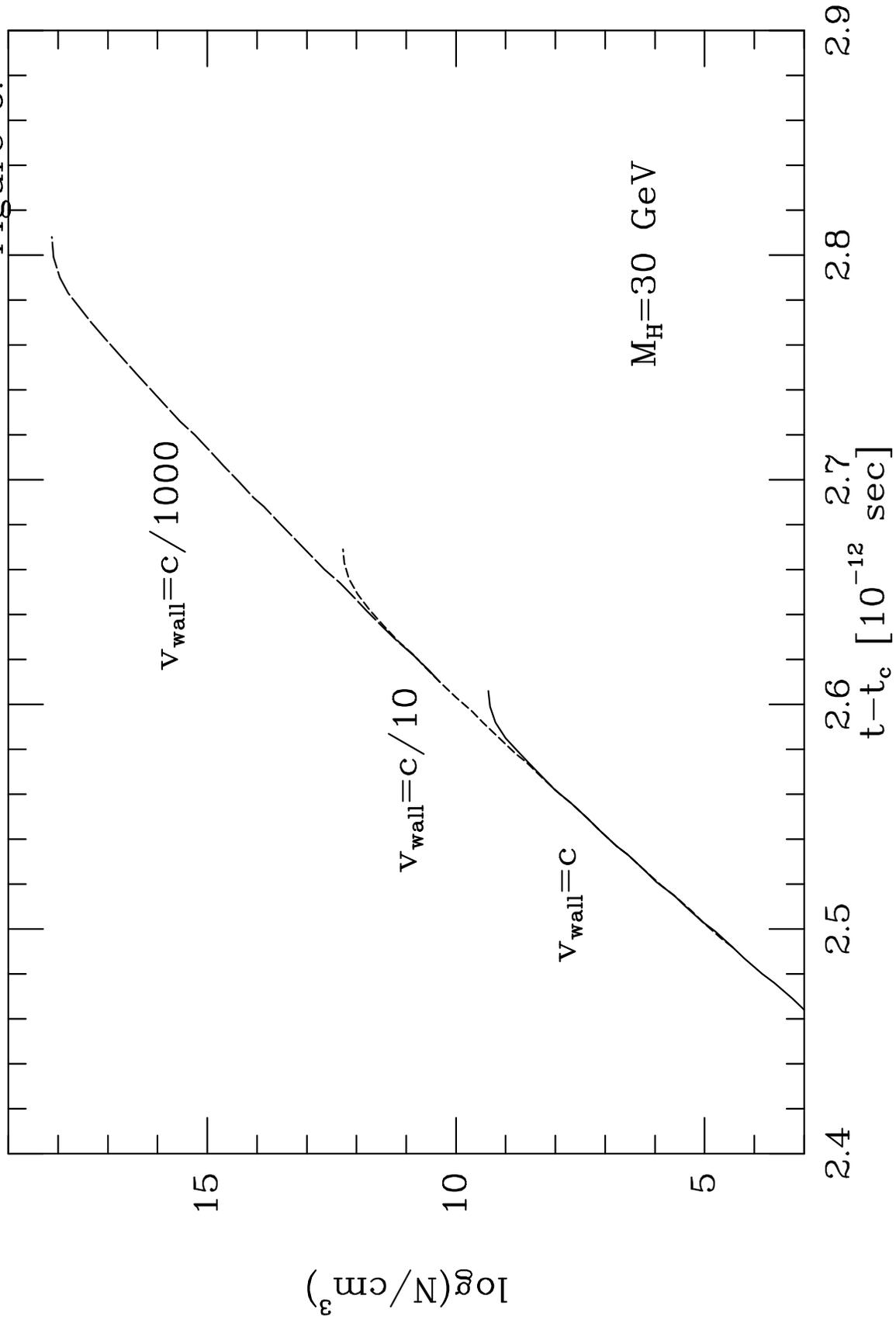
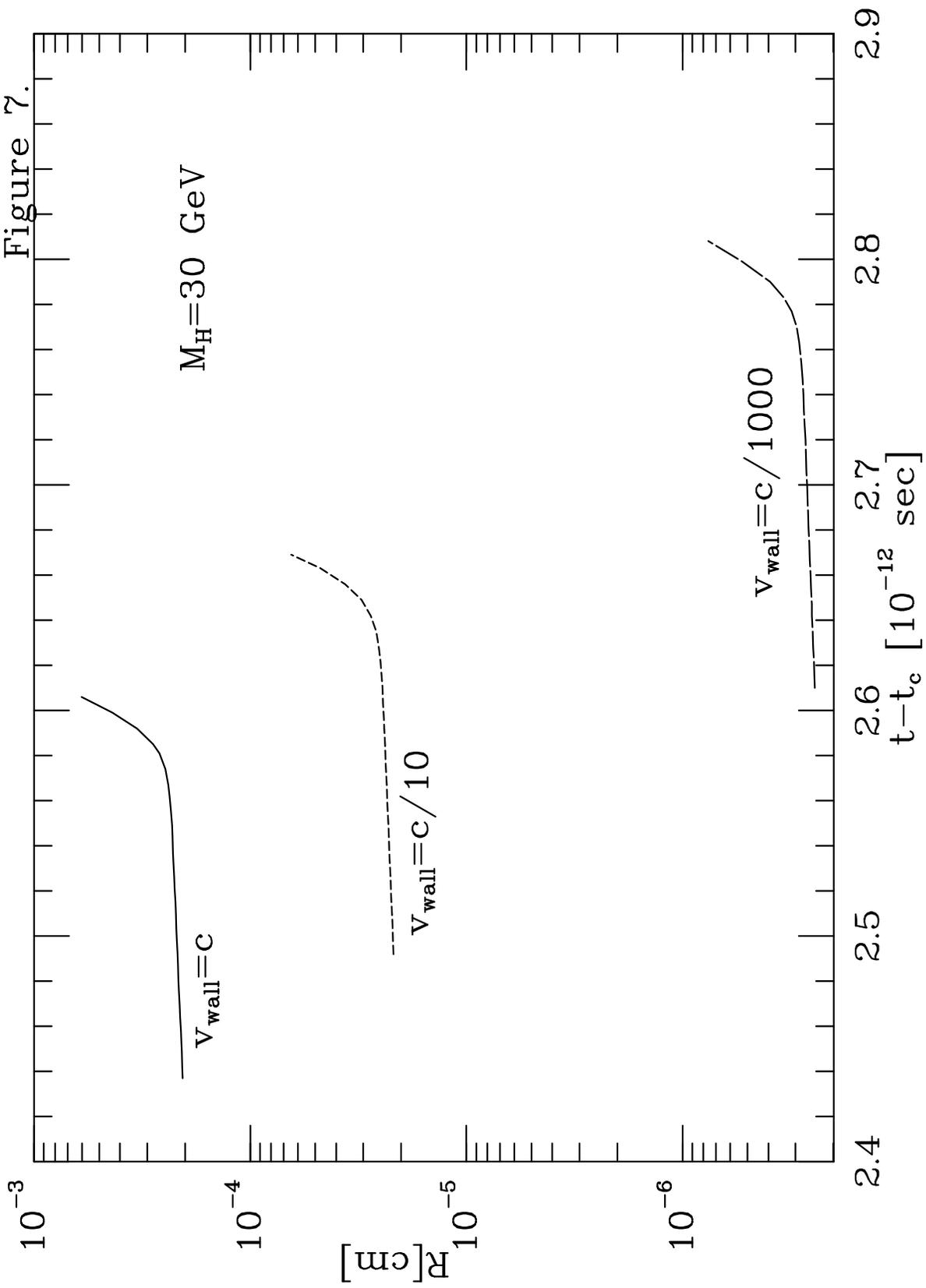
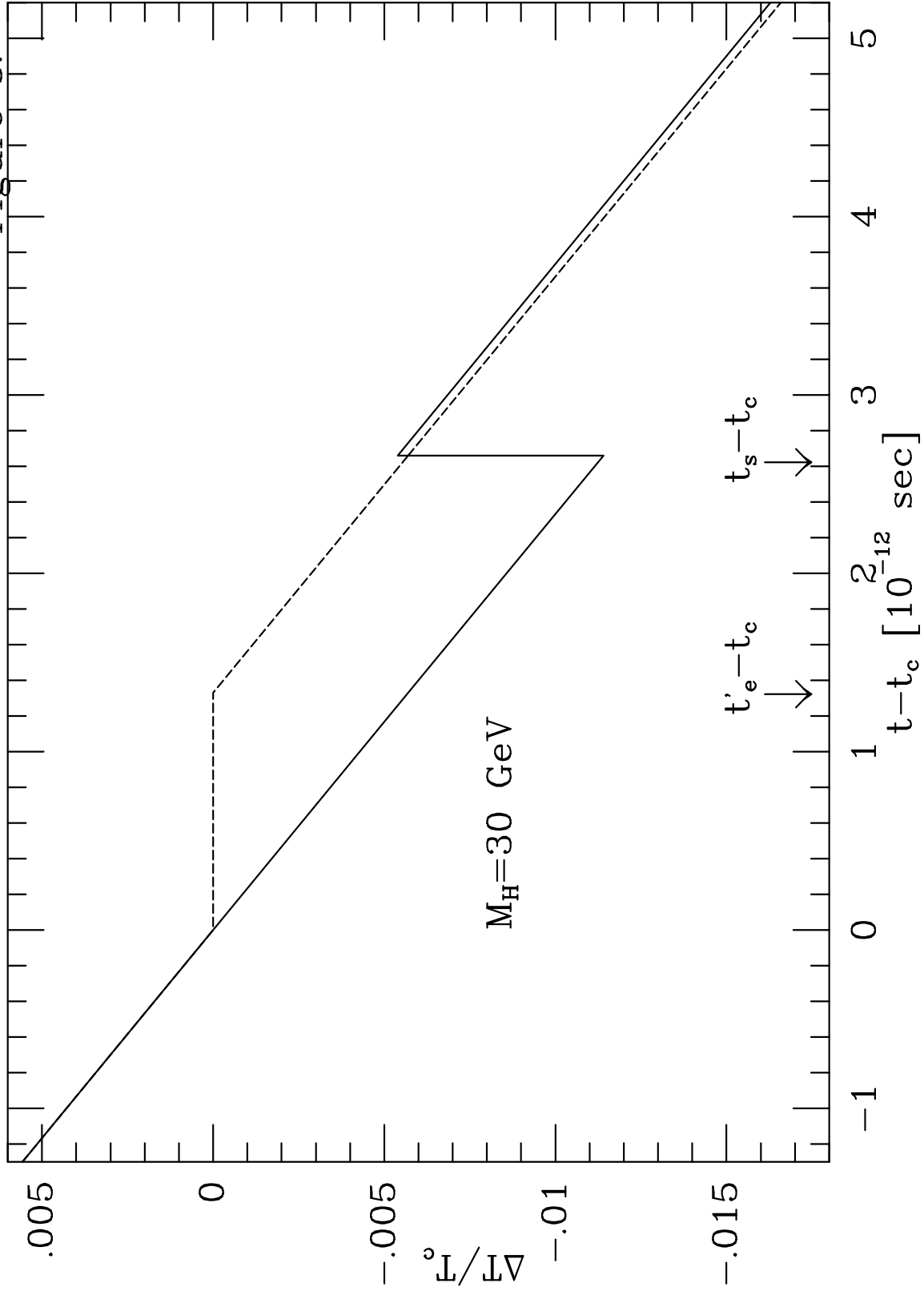
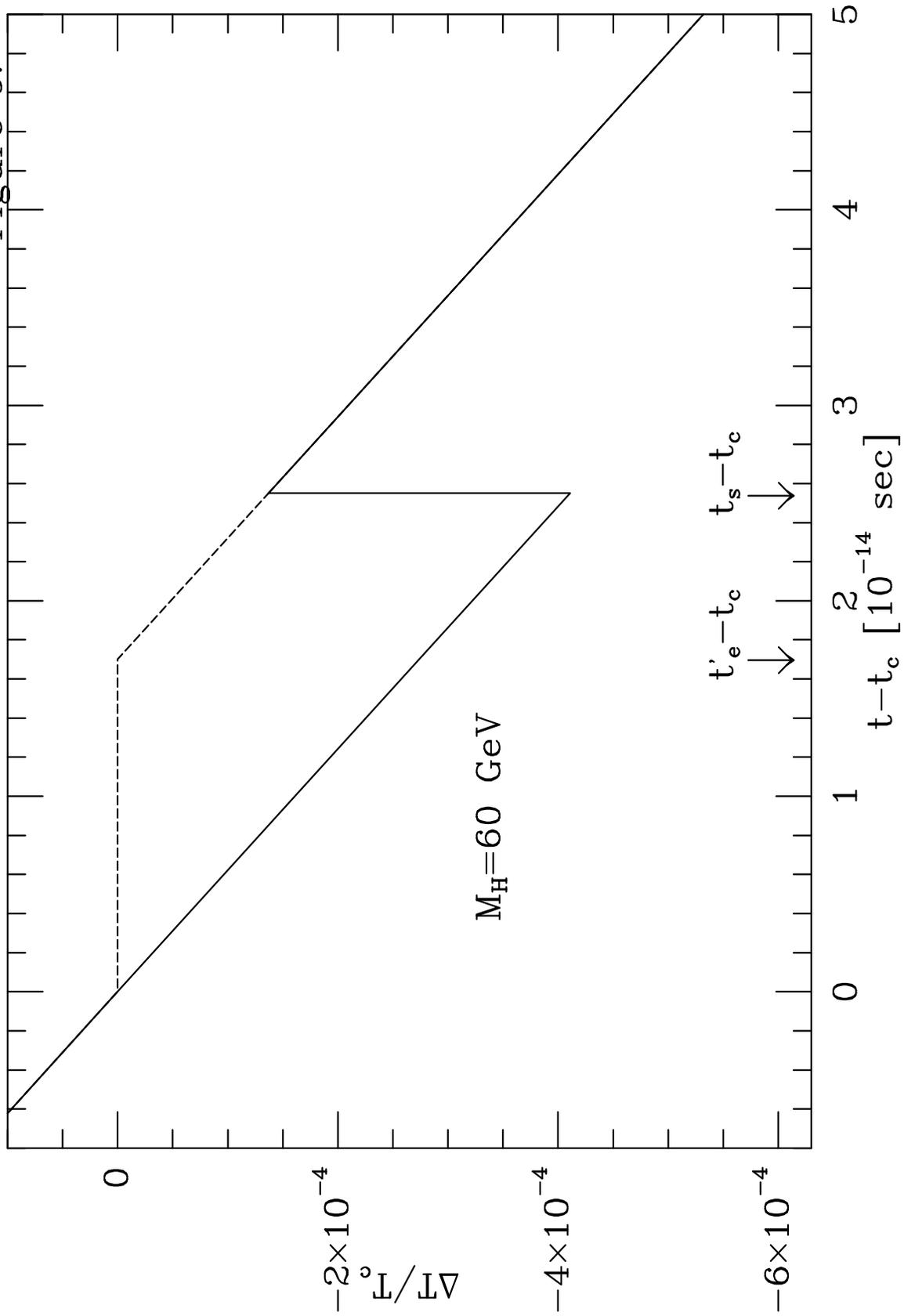
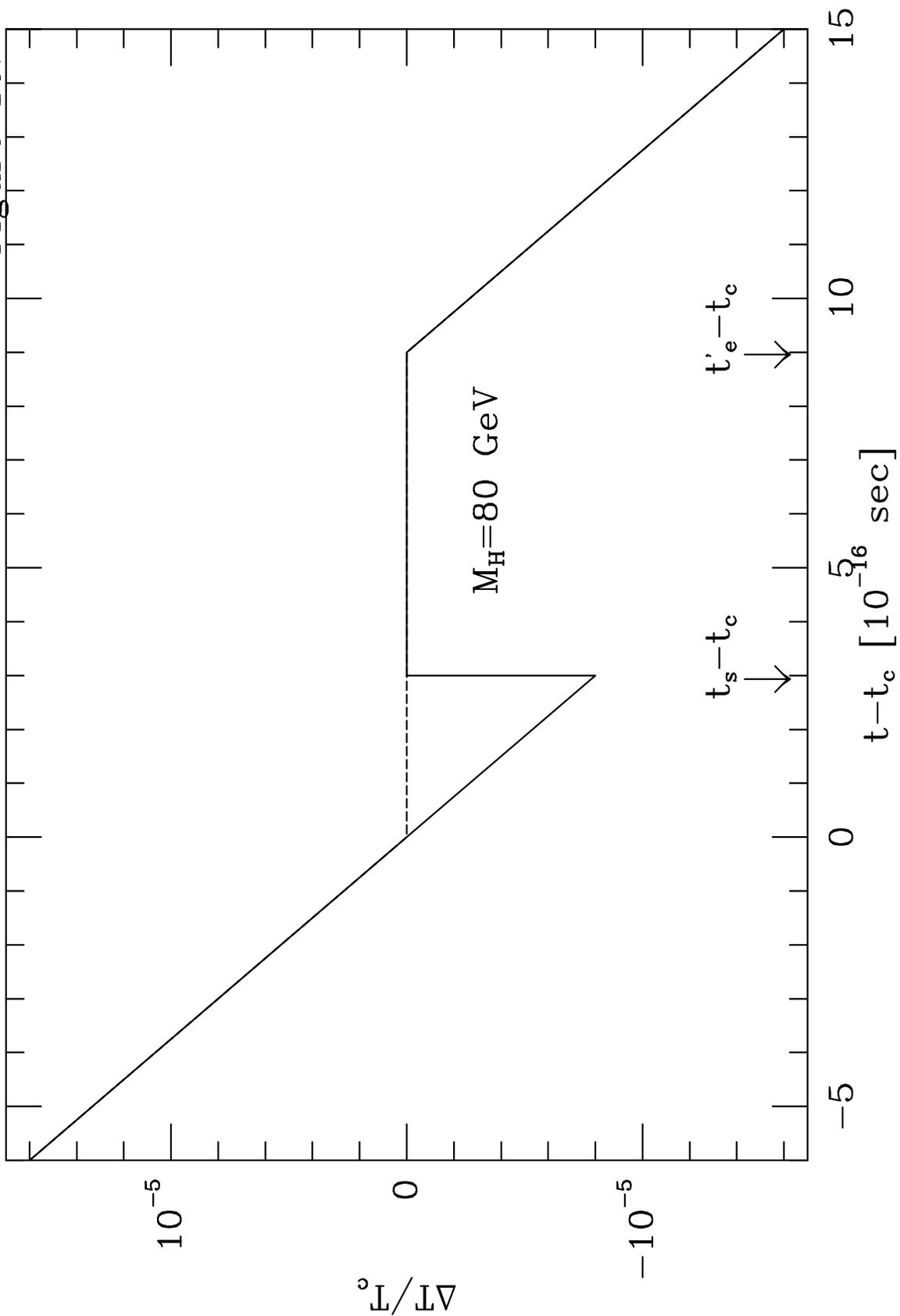
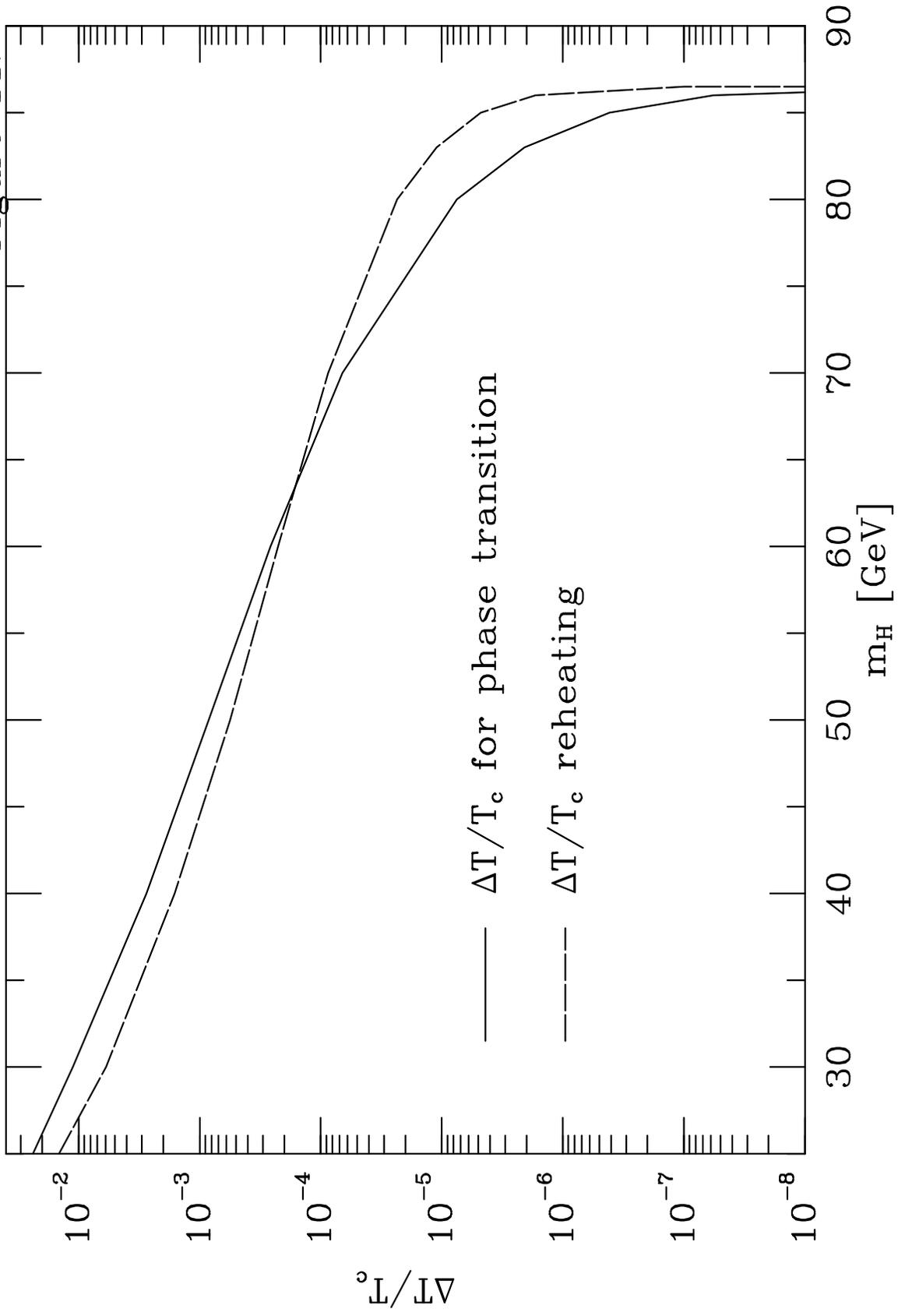
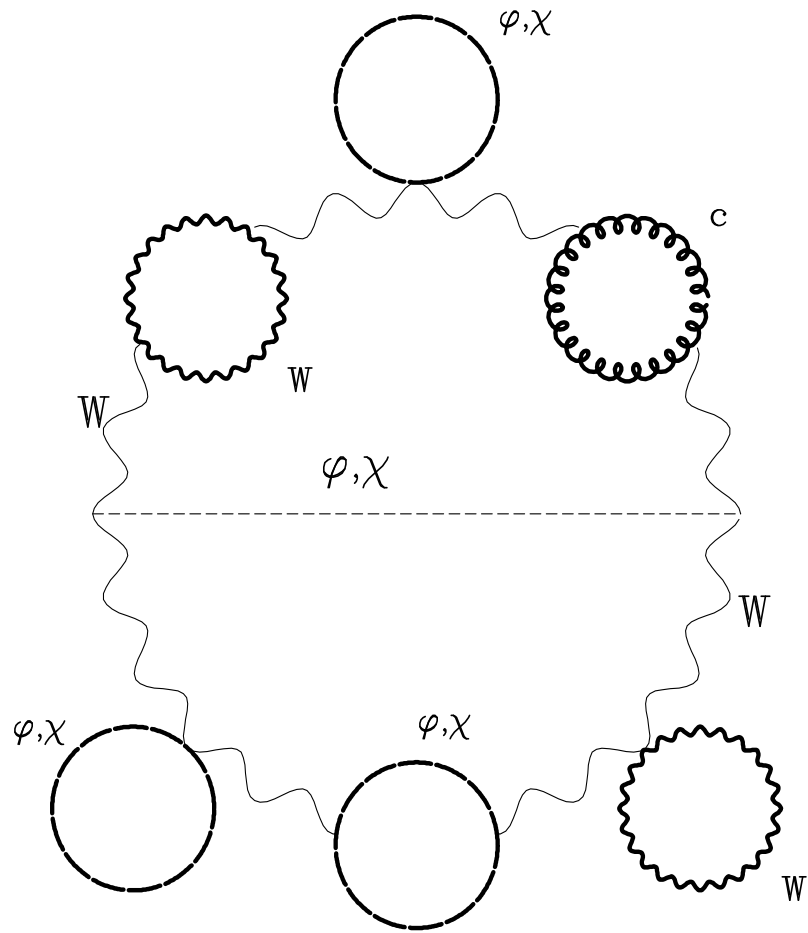
\end{document}